\documentclass[authoryear,final,5p,times,twocolumn]{elsarticle}

\usepackage{amsmath}
\usepackage{graphicx}
\usepackage{float}
\usepackage{natbib}

\usepackage[footnotesize]{caption}
\newcommand{\icar}[1]{}
\newcommand{\prep}[1]{#1}
\prep{\def\figpath{FigsA/}}

\def\Frac#1#2{{{\displaystyle\strut#1}\over{\displaystyle\strut#2}}}

\def\crm{\cr\noalign{\medskip}}
\def\m@th{\mathsurround=0pt}
\def\EQM#1{\vcenter{\normalbaselines\m@th
    \ialign{${\displaystyle ##}$\hfil&&\ ${\displaystyle ##}$\hfil\crcr
    \mathstrut\crcr\noalign{\kern-\baselineskip}
    \noalign{\smallskip}
    #1\crcr\mathstrut\crcr\noalign{\kern-\baselineskip}}}}
\newcommand{\vv}[1]{\boldsymbol{#1}}
\newcommand{\mm}[1]{\tilde{\cal #1}}
\def \be  {\begin{equation}}
\def \ee  {\end{equation}}
\def \ei {\mathrm{e}}
\def \qq {q}
\def \ii {\mathrm{i}}
\def \jj {\mathrm{j}}
\def \kk {\mathrm{k}}
\def \KK {\mathrm{K}}
\def \md {\arrowvert}
\def \ve {\varepsilon}

\def \cQ {\Theta}

\def \cU {{\cal U}}
\def \cV {{\cal V}}
\def \mmu {p}

\def \betal {\beta_{x}}
\def \phil {\phi_{x}}
\def \phis {\psi_{x}}

\def \llabel#1{\label{#1}}

\newcommand\TabA{
\begin{table*}
\begin{center}
\begin{tabular}{|c||c c c c c||c c c c c|} \hline 
$ \quad p \quad $ & $ \phantom{\Frac{1}{16}} $ & $  $ & $ G ( p , e ) $ & $  $ & $  $ 
& $ \phantom{\Frac{1}{16}} $ & $  $ & $ H ( p , e ) $ & $  $ & $  $ \\\hline \hline 
$ \llap{$-$}1   $ & $  $ & $  $ & $ \Frac{9}{4} e^2 $ & $ + $ & $ \Frac{7}{4} e^4 $
  & $ $  & $ $ & $ $ & $  $ & $ \Frac{1}{24} e^4 $ \\ \hline
$ \llap{$-$}1/2 $ & $  $ & $ \Frac{3}{2} e $ & $ + $ & $ \Frac{27}{16} e^3 $ & $ $
  & $ $ & $ $ & $  $ & $ \Frac{1}{48} e^3 $ & $ $ \\ \hline
$  0 $ & $ 1 $ & $ + $ & $ \Frac{3}{2} e^2 $ & $ + $ & $ \Frac{15}{8} e^4 $
  & $ 0 $ & $  $ & $  $ & $ \phantom{\Frac{1}{1}}  $ & $ $ \\ \hline
$  1/2 $ & $  $ & $ \Frac{3}{2} e $ & $ + $ & $ \Frac{27}{16} e^3 $ & $ $
  & $ - $ & $ \Frac{1}{2} e $ & $ + $ & $ \Frac{1}{16} e^3 $ & $ $ \\ \hline
$  1   $ & $  $ & $  $ & $ \Frac{9}{4} e^2 $ & $ + $ & $ \Frac{7}{4} e^4 $
  & $ 1 $ & $ - $ & $ \Frac{5}{2} e^2 $ & $ + $ & $ \Frac{13}{16} e^4 $ \\ \hline
$  3/2 $ & $ $ & $  $ & $  $ & $ \Frac{53}{16} e^3 $ & $ $
  & $ $ & $ \Frac{7}{2} e $ & $ - $ & $ \Frac{123}{16} e^3 $ & $ $ \\ \hline
$  2   $ & $ $ & $ $ & $  $ & $  $ & $ \Frac{77}{16} e^4 $
  & $ $ & $ $ & $ \Frac{17}{2} e^2 $ & $ - $ & $ \Frac{115}{6} e^4 $ \\ \hline
$  5/2 $ & $ $ & $ $ & $ $ & $  $ & $ $
  & $ $ & $ $ & $ $ & $ \Frac{845}{48} e^3 $ & $ $ \\ \hline
$  3   $ & $ $ & $ $ & $ $ & $ $ & $  $
  & $ $ & $ $ & $ $ & $ $ & $ \Frac{533}{16} e^4 $ \\ \hline
\end{tabular} \end{center}
\caption{Coefficients of $ G ( p , e) $ and $ H ( p , e) $ to  $ e^4 $.
The exact expression of these coefficients is given by $  G ( p , e) =
\frac{1}{\pi} \int_0^\pi \left( \frac{a}{r} \right)^3 \exp(\ii \, 2 p M) \, d M
$ and $  H ( p , e) = \frac{1}{\pi} \int_0^\pi \left( \frac{a}{r} \right)^3
\exp(\ii \, 2 \nu) \exp(\ii \, 2 p M) \, d M $. \llabel{TAB1} }
\end{table*}
}

\newcommand\TabB{
\begin{table}
  \begin{center}  
    \begin{tabular}{| c | c c | c c |} \hline
  & \multicolumn{2}{c|}{no CMF} &
  \multicolumn{2}{c|}{$ \nu = 10^{-6} \, \mathrm{m}^2 / \mathrm{s} $} \cr \hline \hline
  $\phantom{--} p \phantom{--}$ & $ \phantom{-} P^-_\mathrm{cap}  \phantom{-}  $ &
   \phantom{-} num.\phantom{--} & $ \phantom{-}  P^-_\mathrm{cap} \phantom{-}
  $ & \phantom{-} num. \phantom{--} \cr \hline
1/2&  1.0 &  1.0  &  29.6 &  31.9   \cr
1/1&  8.5 &  7.9  & 100.0 & 100.0   \cr \hline \hline
  $\phantom{--} p \phantom{--}$ & $ \phantom{-} P^+_\mathrm{cap}  \phantom{-}  $ &
   \phantom{-} num.\phantom{--} & $ \phantom{-}  P^+_\mathrm{cap} \phantom{-}
  $ & \phantom{-} num. \phantom{--} \cr \hline
3/2&  7.7 &  7.2  & 100.0 & 100.0   \cr
2/1&  1.8 &  1.7  & 100.0 & 100.0   \cr
5/2&  0.7 &  1.4  &  45.8 &  46.9   \cr
3/1&  0.3 &  0.4  &  22.3 &  22.3   \cr  
7/2&  0.1 &  0.1  &  11.3 &  11.3   \cr
4/1&  0.1 &   -   &   5.8 &   5.0  \cr
9/2&   -  &   -   &   3.0 &   1.5   \cr
    \hline
    \end{tabular}
  \end{center}
  \caption{Capture probabilities in several spin-orbit resonances (in percent). 
  The first column ($ P^\pm_\mathrm{cap} $) refers to the theoretical estimation
  given by expression (\ref{040820i}), while the next column (num.) refers to
  the estimation obtained running a numerical simulation with 2000 different
  initial conditions. We used $ (B-A)/C_m = 1.2 \times 10^{-4} $, and a constant
  eccentricity $ e = 0.206 $. \llabel{TBNat3}}    
\end{table}
}
\newcommand\TabC{
\begin{table*}
  \begin{center}
    \begin{tabular}{| c | c c c c c c c c c|}
    \hline
  $\ve_0$  & \phantom{--}1/2 &
  \phantom{--}1/1 & \phantom{--}3/2 & \phantom{--}2/1 & \phantom{--}5/2 &
  \phantom{--}3/1 & \phantom{--}7/2 & \phantom{--}4/1 & \phantom{--}9/2 \cr \hline
  $0^\circ$&  - & - &	-  & 34.3 & 30.3 & 18.6 & 10.6 & 4.9 & 1.5 \cr
  $5^\circ$&  - & - &	-  & 32.8 & 31.1 & 19.1 & 10.8 & 4.9 & 1.5 \cr
 $10^\circ$&  - & - &  1.0 & 32.2 & 28.6 & 20.3 & 10.8 & 5.6 & 1.5 \cr
 $15^\circ$&  - & - &  2.5 & 30.8 & 29.8 & 19.8 & 10.7 & 5.3 & 1.3 \cr
 $30^\circ$&  - & - &  3.9 & 32.1 & 29.5 & 17.9 & 11.3 & 4.3 & 1.0 \cr
 $45^\circ$&  - & - &  1.7 & 34.8 & 27.9 & 19.1 & 10.9 & 5.1 & 0.8 \cr
 $60^\circ$&  - & - &  5.0 & 32.0 & 28.1 & 20.0 &  9.2 & 5.2 & 0.8 \cr
 $75^\circ$&  - & - &  7.4 & 29.9 & 28.9 & 18.6 &  9.5 & 4.6 & 1.3 \cr
 $90^\circ$&  - & - &  2.4 & 31.9 & 29.4 & 19.2 & 10.7 & 5.5 & 1.0 \cr
$105^\circ$&  - & - &  5.3 & 33.8 & 27.5 & 18.7 &  9.0 & 4.9 & 1.0 \cr
$120^\circ$&  - & - &  4.2 & 32.9 & 29.4 & 18.5 & 10.2 & 3.9 & 1.1 \cr
$135^\circ$&  - & - &  3.3 & 34.7 & 28.2 & 19.0 &  9.2 & 5.0 & 0.8 \cr
$150^\circ$&  - & - &  2.5 & 36.3 & 29.7 & 18.1 &  9.9 & 3.3 & 0.4 \cr
$165^\circ$&  - & - & 23.6 & 46.6 & 23.3 &  4.6 &  1.9 &  -  &  -  \cr
$170^\circ$&  - & 25.5& 59.2 & 11.7 &  3.2 &  0.5 &   -  &  -  &  -  \cr
$175^\circ$&  2.4 &  6.8 & 6.0 & 0.8 &  -  & - & - & - & - \cr
$175^\circ (*)$& 28.2 & 55.9 & - & - &  -  & - & - & - & - \cr
$180^\circ (*)$& 31.9 & 68.1 & - & - & - & - & - & - & - \cr
    \hline
    \end{tabular} 
  \end{center}
  \caption{Capture probabilities (in percent) in several spin-orbit resonances for different
  initial obliquities. Tidal effects and full core-mantle friction are included
  with $ \nu = 10^{-6} \, \mathrm{m}^2 \mathrm{s}^{-1} $. We performed a
  numerical simulation with 2000 close initial conditions with $ (B-A)/C_m = 1.2
  \times 10^{-4}$ and $ e = 0.206 $. We see that crossing a resonance with an
  obliquity different from zero substantially modifies the chances of being
  captured in a specific spin-orbit resonance.
  $(*)$: these final equilibria are achieved for a
  final obliquity of $180^\circ$ and a negative rotation rate.
  \llabel{TM01}}  
\end{table*}
}

\newcommand\FigA{
\begin{figure}[h!]
  \begin{center}
    \includegraphics*[width=8cm]{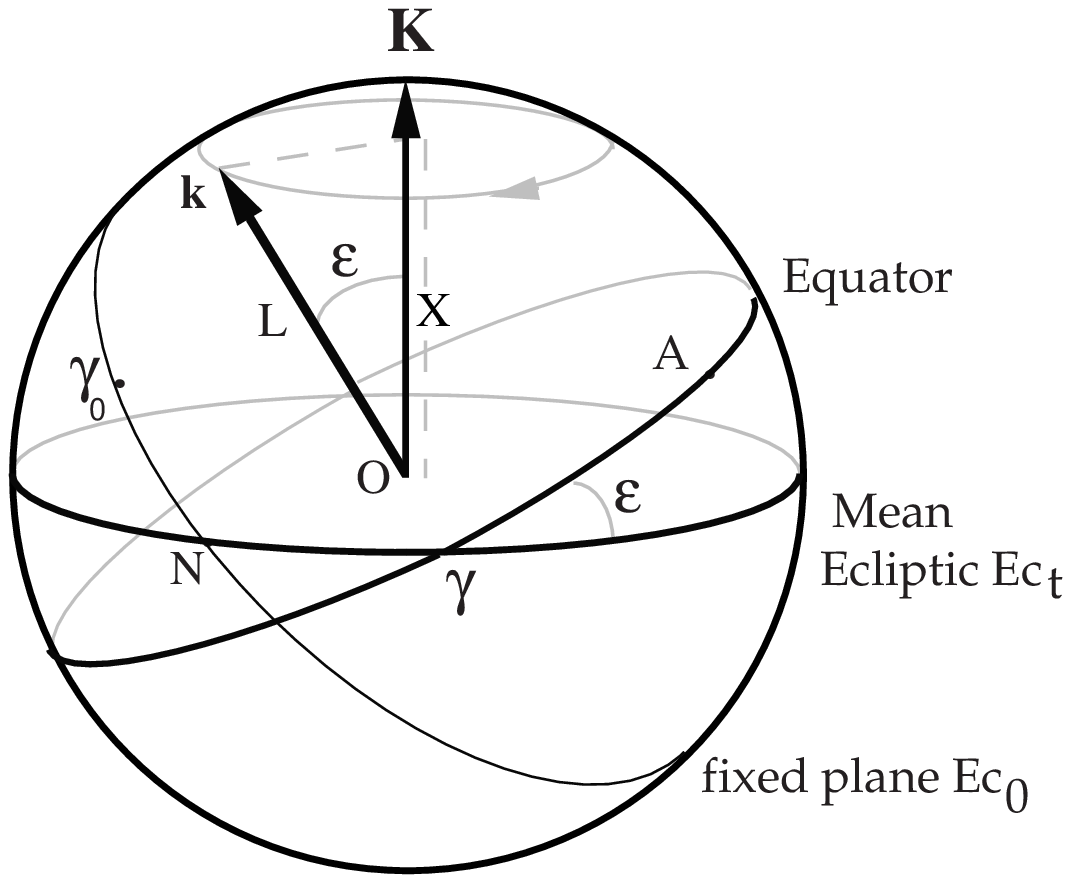}
    \caption{Andoyer's canonical variables. $ L $ is the projection of the total
    rotational angular momentum vector $ \vv{L} $ on the principal axis of
    inertia $ \vv{\kk}
    $ and $ X $ its projection on the normal to the ecliptic $ \vv{\KK} $. The
    angle between the equinox of date $ \gamma $ and a fixed point of the
    equator $ A $ is the hour angle $ \theta $, and $ \psi = \gamma $ N + N $
    \gamma_0 $ is the general precession angle. The direction of $ \gamma_0 $ is on
    a fixed plane $ {E_c}_0 $, while $ \gamma $ is on the mean orbital (or
    ecliptic) $ {E_c}_t $ of date $ t $.
   \llabel{Fig01} }
  \end{center}
\end{figure}
}

\newcommand\FigB{
\begin{figure}[h!]
  \begin{center}
     \includegraphics*[width=8cm]{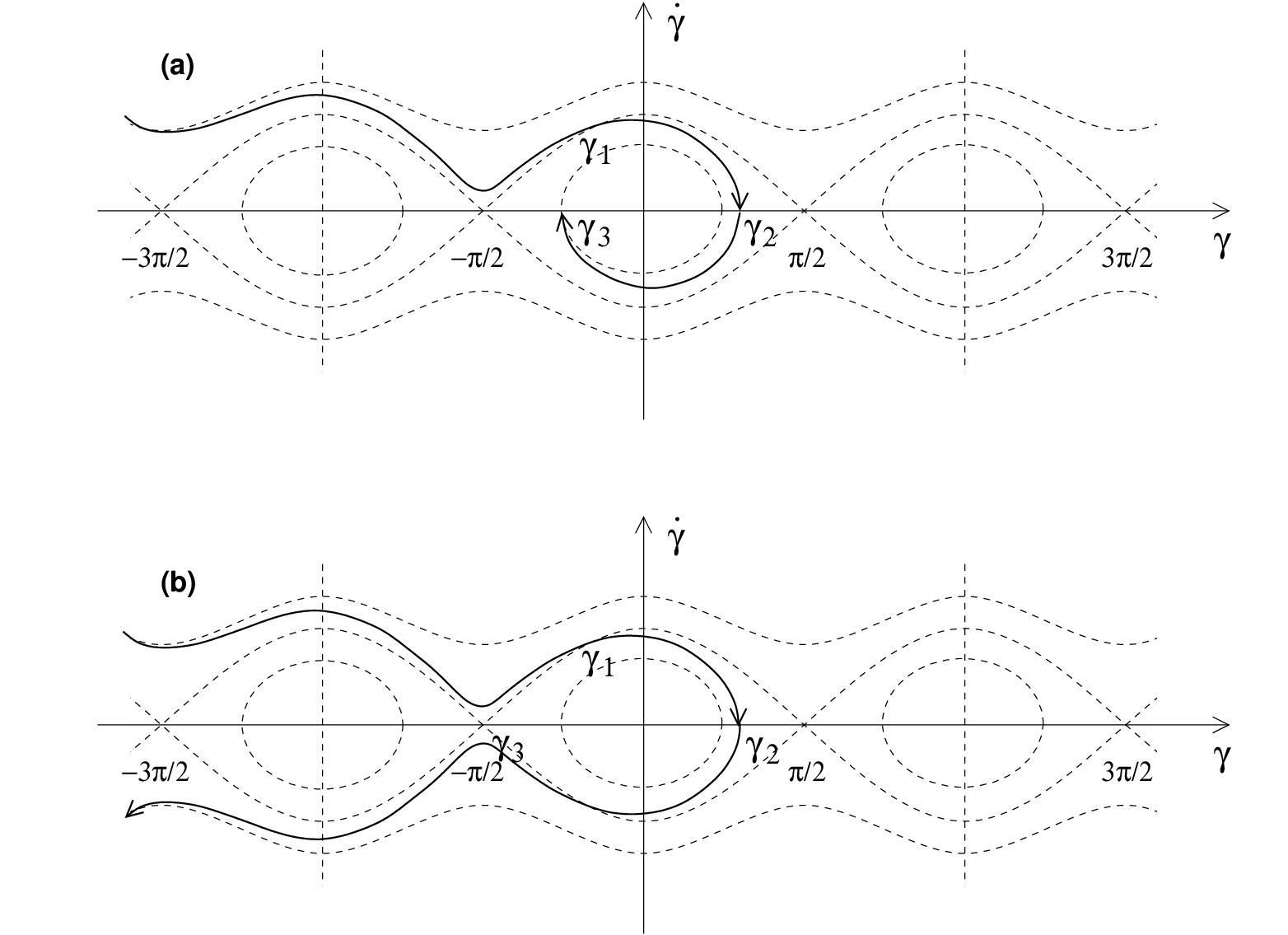}
   \caption{Capture in the $ \gamma = \theta - p M - \phil $ resonance in the phase
   space ($ \gamma $,$ \dot \gamma $). In the first situation (a) the planet is
   captured, while in the second one (b) it manages to cross the resonance
   without being trapped. \llabel{Fig02} }
  \end{center}
\end{figure}
}

\newcommand\FigC{
\begin{figure}[h!]
  \begin{center}
    \includegraphics*[width=8cm]{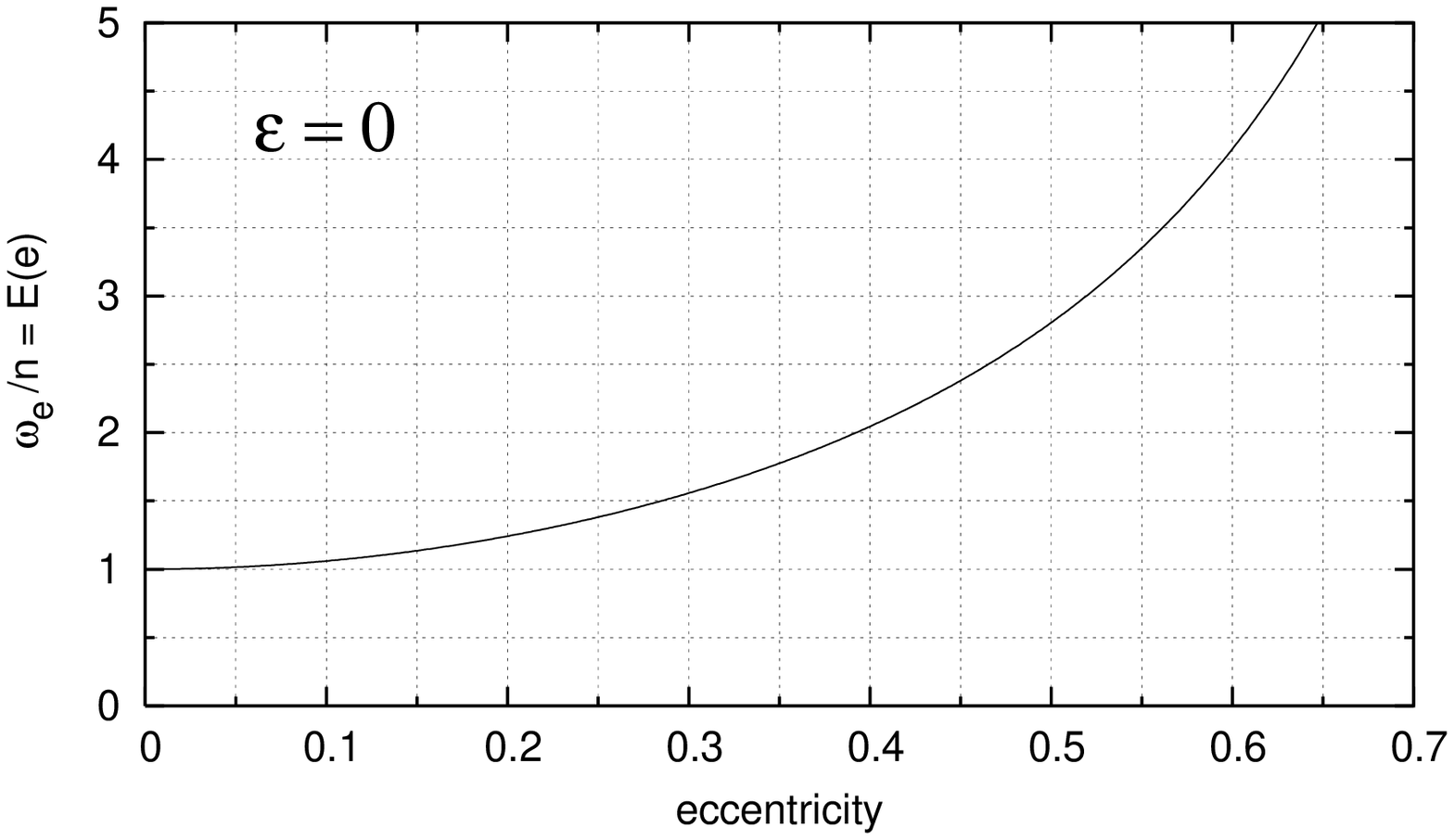}
  \caption{Evolution of the equilibrium rotation rate $ \omega_e / n = E(e) $ with the
  eccentricity when $ \ve = 0^\circ $ using the viscous model. 
  As the eccentricity increases, $ \omega_e $ also increases. The tidal effects
  lead the planet to exact resonance when the eccentricity is respectively
  $ e_{1/1} = 0 $, $ e_{3/2} = 0.284926803 $ and $ e_{2/1} = 0.392363112 $.
   \llabel{Fig04} }
  \end{center}
\end{figure}
}

\newcommand\FigD{
\prep{\begin{figure*}}
\icar{\begin{figure}}
  \begin{center}
    \includegraphics*[width=16cm]{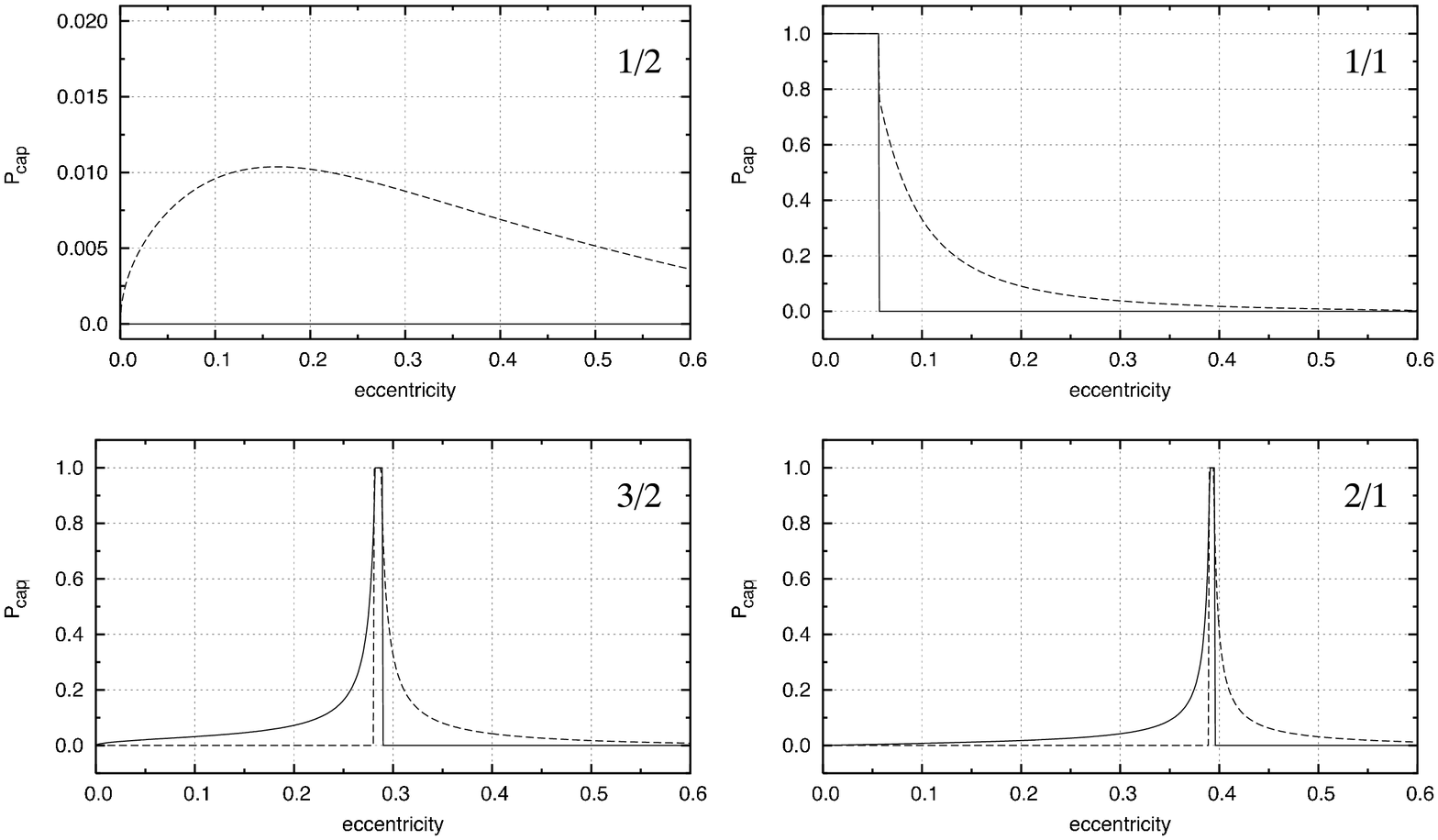} 
  \caption{Probability of capture in some spin-orbit resonances for eccentric
  orbits with $ \ve = 0^\circ $ and $ (B-A)/C_m = 1.2 \times 10^{-4} $, when
  using the viscous model for tides and no core-mantle friction.  
  The dashed line corresponds to a planet increasing its spin 
  from slower rotation rates, while the solid line corresponds to a planet
  de-spinning from faster rotation rates. In this last situation, for all
  resonances but the 1/2, as the eccentricity increases, capture
  probability in higher-order resonances increases. However, it suddenly
  decays to zero when the equilibrium rotation rate falls outside the resonance
  width. For the 1/2 resonance, the planet can only be captured when increasing
  its spin from lower values and with low probability (the maximum being about
  1.04\% for $ e = 0.165 $). \llabel{Fig05} }
  \end{center}
\prep{\end{figure*}}
\icar{\end{figure}}
}

\newcommand\FigE{
\prep{\begin{figure*}}
\icar{\begin{figure}}
  \begin{center}
    \includegraphics*[width=16cm]{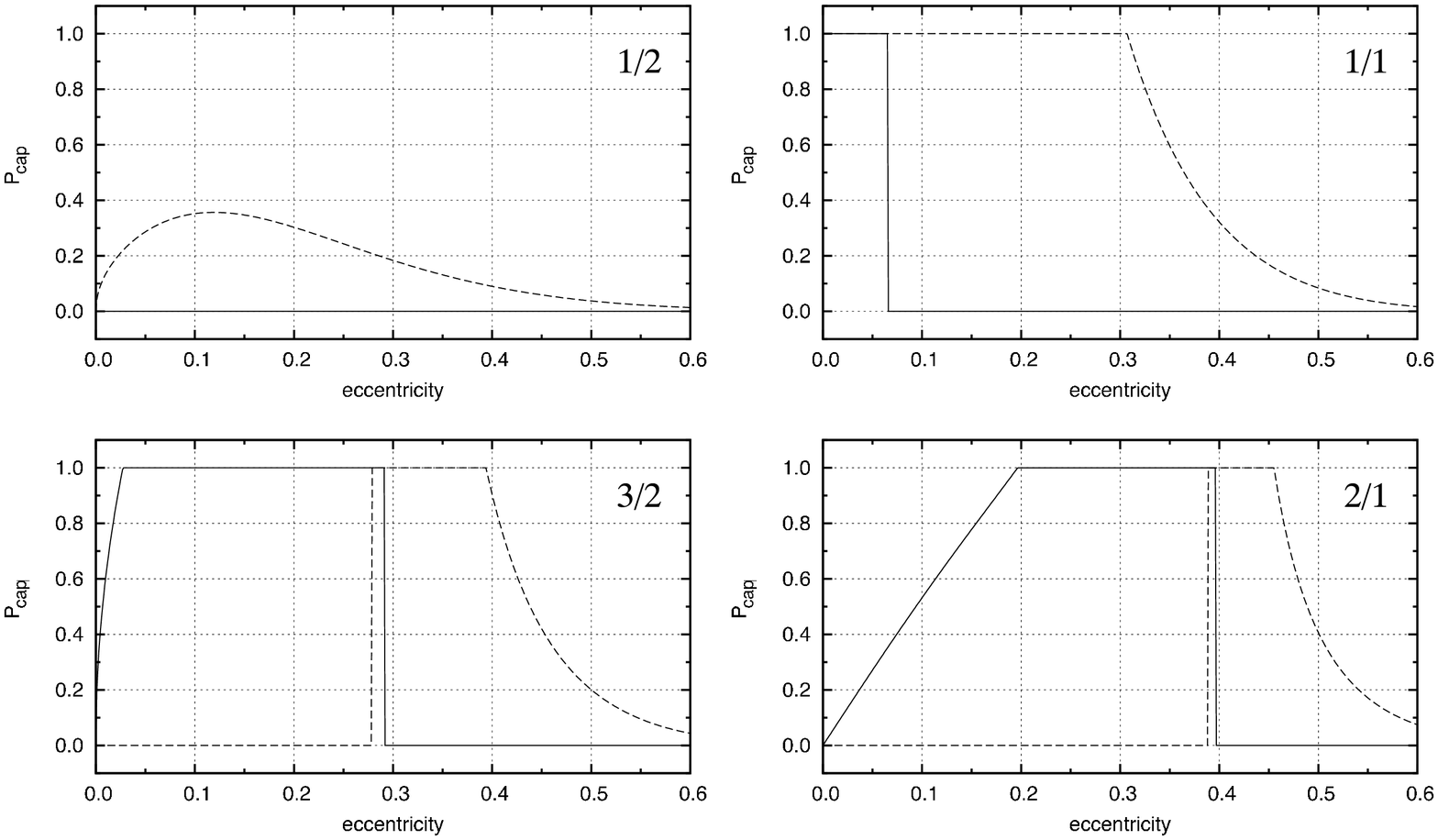} 
  \caption{Probability of capture in some spin-orbit resonances for eccentric
  orbits with $ \ve = 0^\circ $ and $ (B-A)/C_m = 1.2 \times 10^{-4} $, when
  using the viscous model for tides and $ \nu = 10^{-6} \, \mathrm{m}^2
  \mathrm{s}^{-1} $.  
  The dashed line corresponds to a planet increasing its spin
  from slower rotation rates, while the solid line corresponds to a planet
  de-spinning from faster rotation rates. 
  When comparing with figure~\ref{Fig05} (absence of core-mantle friction), we
  mainly notice that the eccentricity values that provide 100\% of chances of
  being captured largely increase.
  For the 1/2 resonance, the planet can only still be captured when increasing
  its spin from lower values but now also with higher chances (the maximum being about
  36\% for $ e = 0.12 $). \llabel{Fig05B} }
  \end{center}
\prep{\end{figure*}}
\icar{\end{figure}}
}

\newcommand\FigF{
\begin{figure}[h!]
  \begin{center}
    \begin{tabular}{c}
    \includegraphics*[width=8cm]{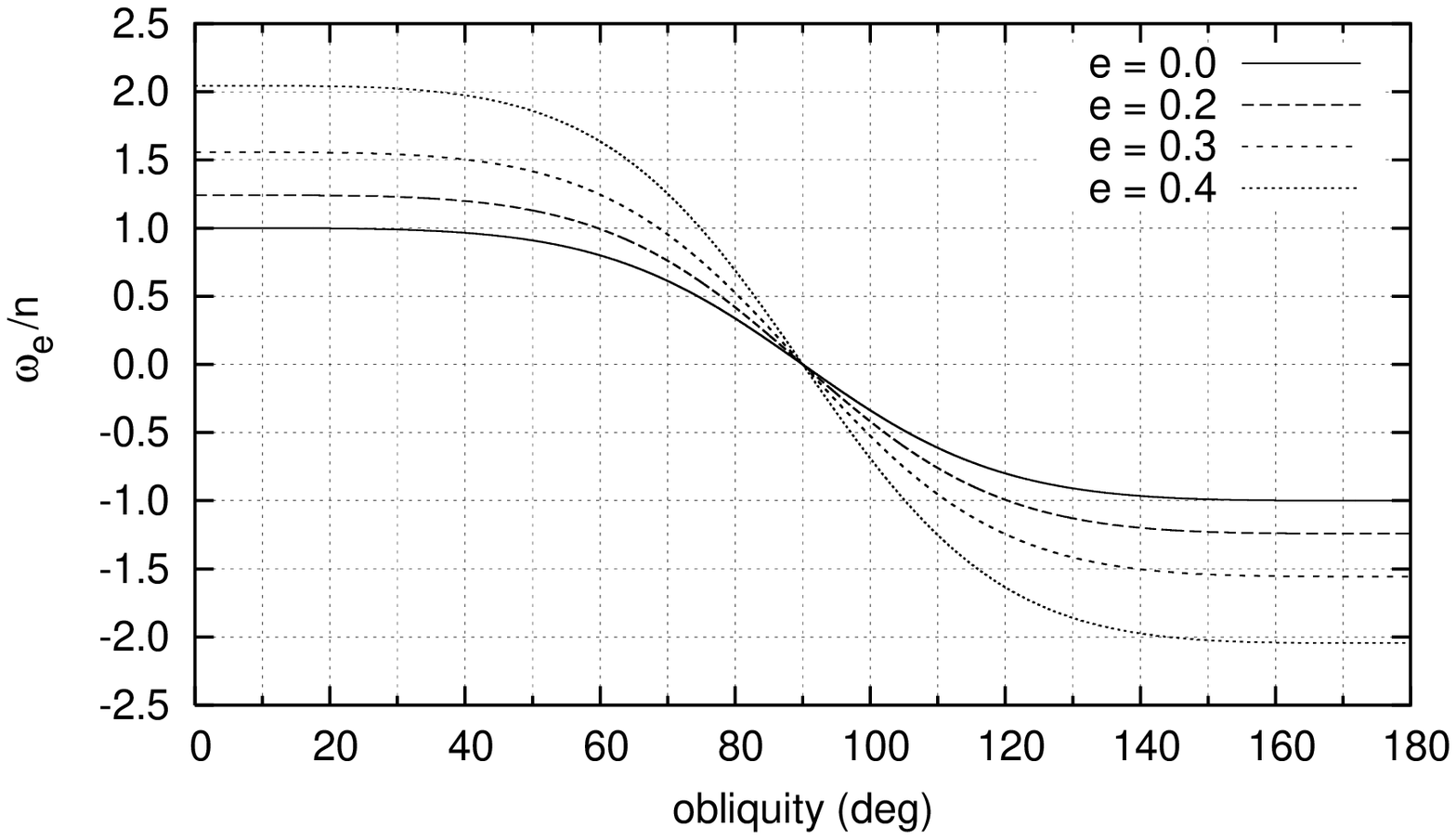}
  \end{tabular}
  \caption{Evolution of the equilibrium rotation rate $ \omega_e $ with the
  obliquity for fixed eccentricities and using the viscous tidal model. 
  The equilibrium rotation rate decreases as the obliquity increases.
  For $ \ve = 90^\circ $ we have $ \omega_e = 0 $ for all eccentricities. 
  Although we have $ \omega_e < 0 $ for $ \ve > 90^\circ $, notice that the
  equilibrium spin still corresponds to a prograde rotation state.
  \llabel{Fig06} }
  \end{center}
\end{figure}
}

\newcommand\FigH{
\begin{figure}[h!]
  \begin{center}
    \includegraphics*[width=8.cm]{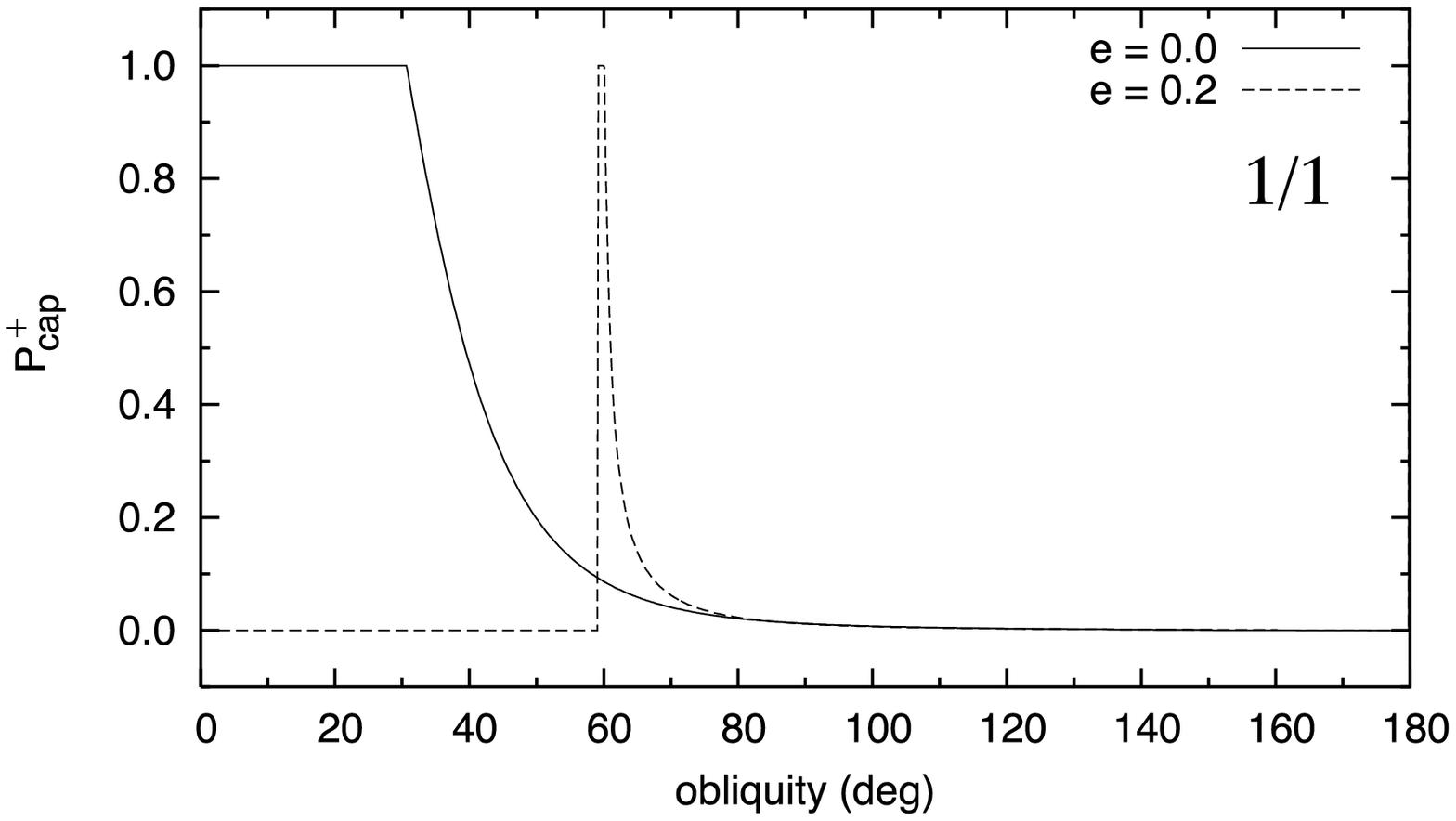} 
  \caption{Probability of capture in the 1/1 spin-orbit resonance when
  de-spinning from faster rotation rates for different obliquities and $ e = 0 $
  (solid line) or $ e = 0.2 $ (dashed line). We used the viscous tidal model and  $
  (B-A)/C_m = 1.2 \times 10^{-4} $. We see that the capture probability decreases as
  the obliquity increases. Therefore, unless $ \ve < 30.66^\circ $, the
  synchronous resonance is no longer the last possible stage for the spin
  evolution.  
  \llabel{Fig07} }
  \end{center}
\end{figure}
}

\newcommand\FigI{
\begin{figure}
  \begin{center}
    \includegraphics*[width=8cm]{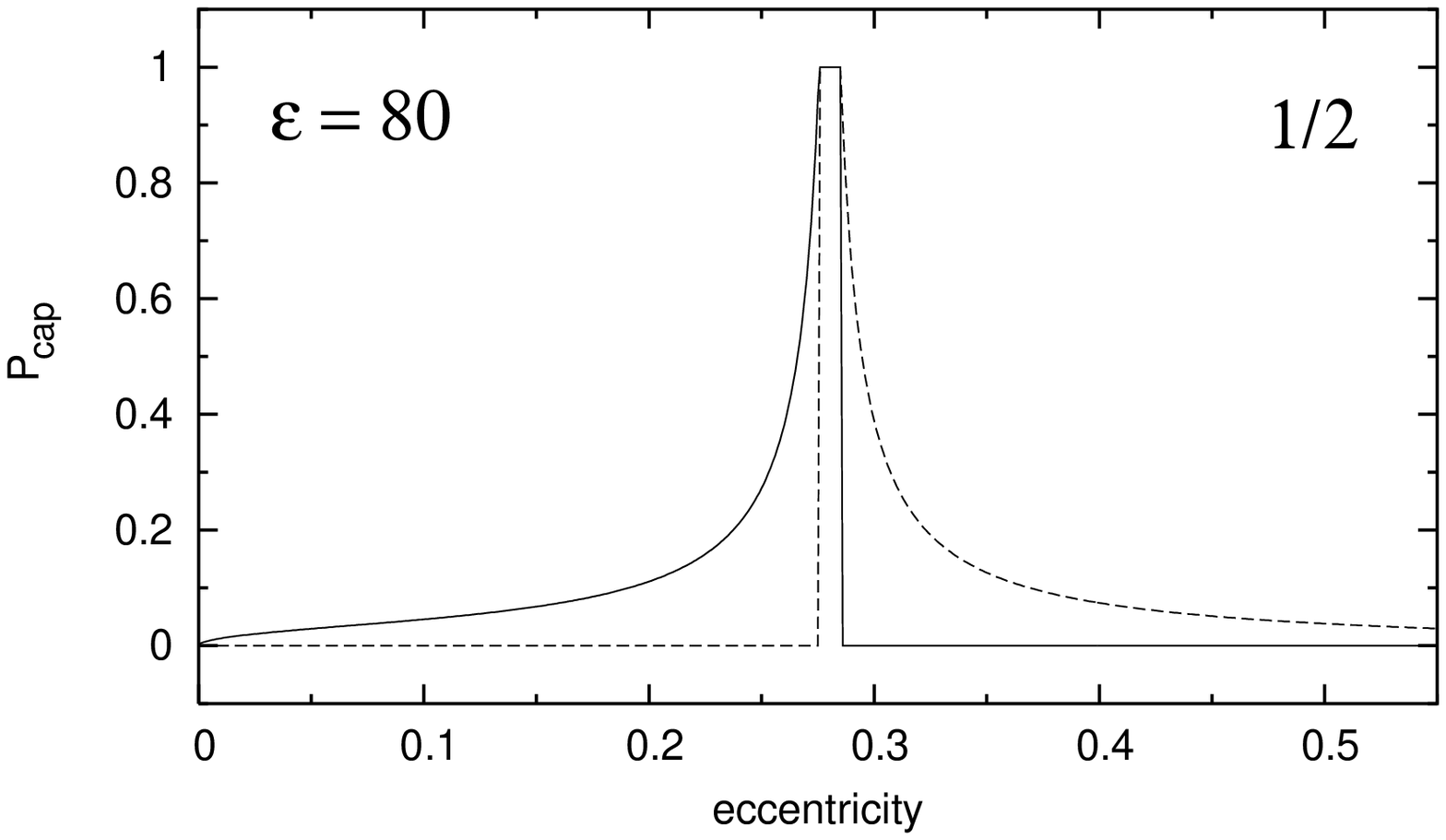} 
  \caption{Probability of capture in the 1/2 spin-orbit resonance for eccentric
  orbits when $ \ve = 80^\circ $, using the viscous tidal model and $ (B-A)/C_m = 1.2
  \times 10^{-4} $. The solid line corresponds to a planet de-spinning from
  faster rotation rates, while the dashed line corresponds to a planet
  increasing its spin from slower rotation rates. According to expression
  (\ref{040819a}), for $ \ve > 74.5^\circ $ the equilibrium rotation rate $
  \omega_e $ is below the 1/2 resonance and for those obliquities $
  P_\mathrm{cap} \ne 0 $ (Eq.\ref{040819b}).   \llabel{Fig09} }
  \end{center}
\end{figure}
}

\newcommand\FigJ{
\begin{figure}
  \begin{center}
    \includegraphics*[width=8cm]{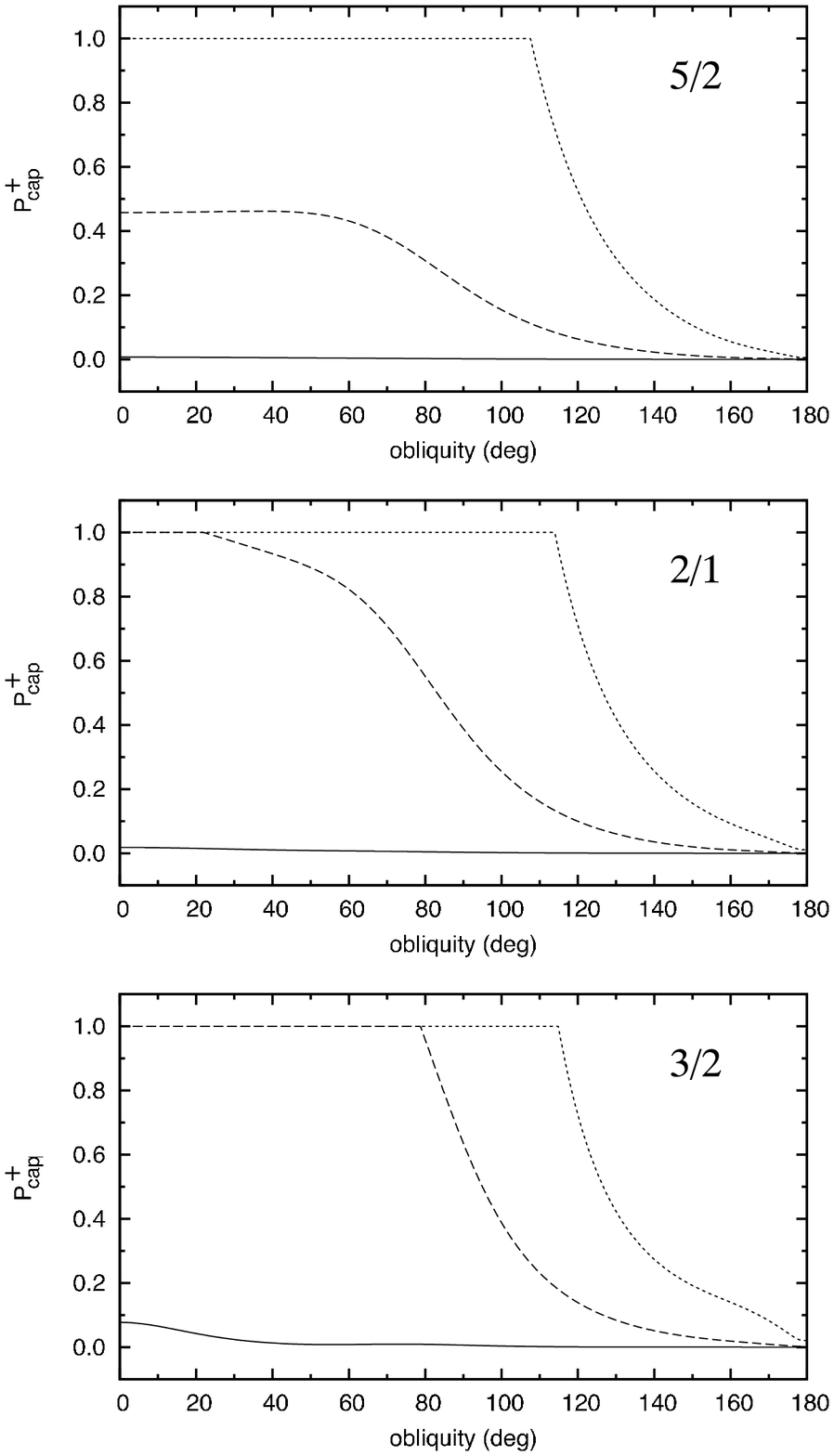}
  \caption{Probability of capture in some spin-orbit resonances when
  de-spinning from faster rotation rates for different
  obliquities when $ e = 0.206 $ and $ (B-A)/C_m = 1.2 \times 10^{-4} $. 
  We compute the probability when tidal effects are
  considered alone (solid line) and when core-mantle friction is also included,
  using two different effective viscosities: $ \nu = 10^{-6} \, \mathrm{m}^2
  \mathrm{s}^{-1} $ (dashed line) and $ \nu = 10^{-4} \, \mathrm{m}^2
  \mathrm{s}^{-1} $ (dotted line).
  Although the resonant core-mantle coupling greatly increases the chances of
capture in resonance at zero obliquity ($ P_\mathrm{cap}^+ = 100$\% for the 2/1
and 3/2 resonances), for high values of the obliquity this probability
considerably decreases, allowing the rotation of the planet to evolve into 
lower-order equilibrium configurations.
  \llabel{Fig08} } 
  \end{center}
\end{figure}
}

\newcommand\FigK{
\begin{figure}
  \begin{center}
    \includegraphics*[height=8cm,angle=90]{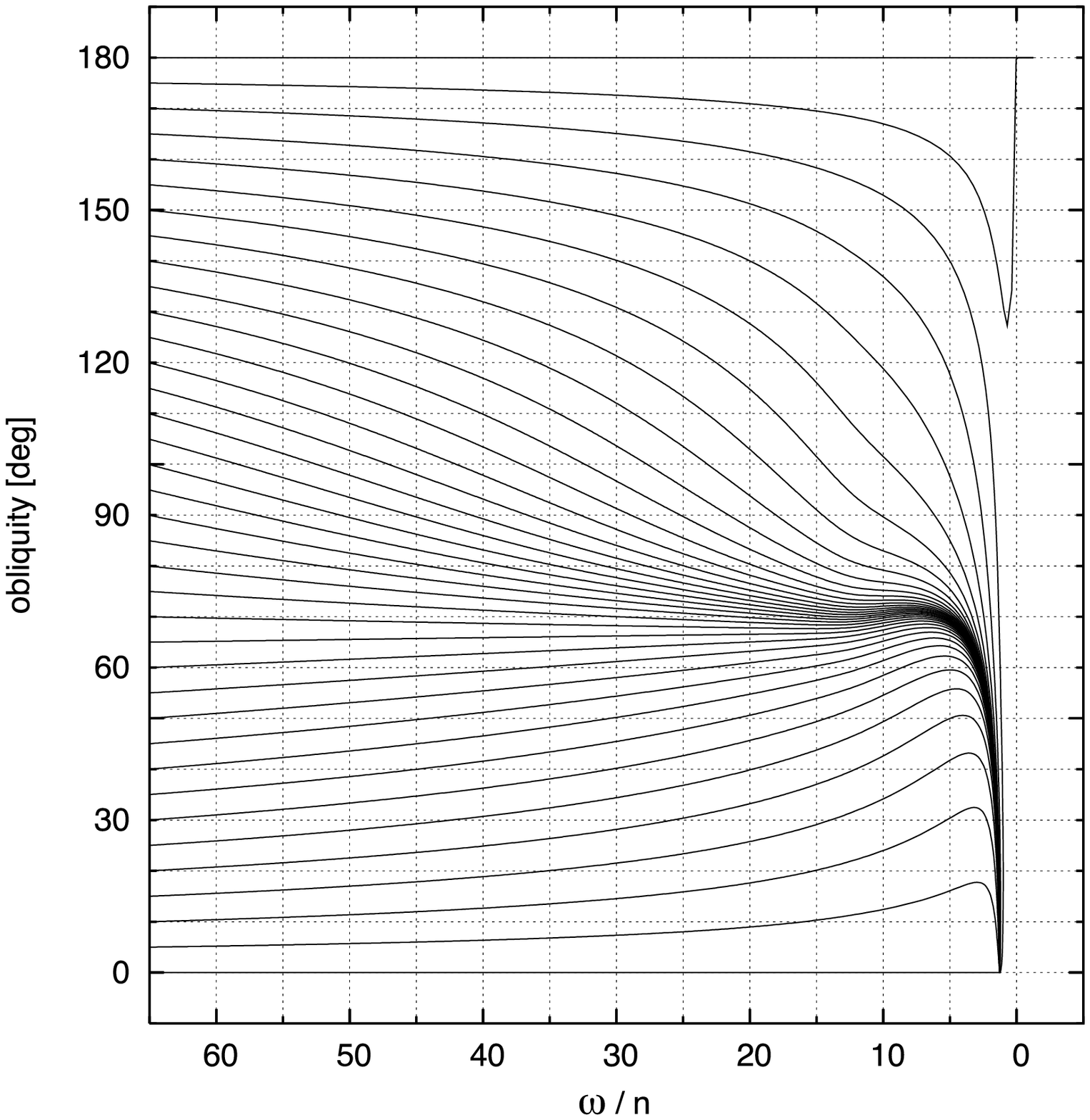}
  \caption{Obliquity evolution with the rotation rate for an initial
  rotation period of about 32~h ($ \omega_i = 65 \, n $). We observe two distinct
  behaviors: for initial obliquities smaller than $ 175^\circ $ the obliquity is
  brought to zero. For higher initial obliquities the final obliquity is $
  180^\circ $ and the final rotation rate is negative. \llabel{FM01}}
  \end{center}
\end{figure}
}

\begin{document}

\begin{frontmatter}

\title{Long-term evolution of the spin of Mercury \\ I. Effect of the
obliquity and core-mantle friction}

\author[lab1,lab2]{Alexandre C. M. Correia}
\author[lab2]{Jacques Laskar}

\address[lab1]{Departamento de F\'\i sica, Universidade de Aveiro,
Campus de Santiago, 3810-193 Aveiro, Portugal}

\address[lab2]{Astronomie et Syst\`emes Dynamiques, IMCCE-CNRS UMR8028, 
Observatoire de Paris, UPMC, 77 Av. Denfert-Rochereau, 75014 Paris, France}

\begin{abstract}

The present obliquity of Mercury is very low (less than $ 0.1^\circ $), which led
previous studies to always adopt a nearly zero obliquity during the planet's
past evolution.
However, the initial orientation of Mercury's rotation axis is unknown and
probably much different than today.
As a consequence, we believe that the obliquity could have been significant when
the rotation rate of the planet first encountered spin-orbit resonances.
In order to compute the capture probabilities in resonance for any evolutionary
scenario, we present in full detail the dynamical equations governing the
long term evolution of the spin, including the obliquity contribution.

The secular spin evolution of Mercury results from tidal interactions with the
Sun, but also from viscous friction at the core-mantle boundary.
Here, this effect is also regarded with particular attention.
Previous studies show that a liquid core enhances
drastically the chances of capture in spin-orbit resonances.
We confirm these results for null obliquity, but we find that the capture
probability generally decreases as the obliquity increases.
We finally show that, when core-mantle friction is combined with obliquity
evolution, the spin can evolve into some unexpected configurations as the
synchronous or the 1/2 spin-orbit resonance. 

\end{abstract}


\end{frontmatter}

\section{Introduction}

The present rotation rate of Mercury was discovered by \citet{Pettengill_Dyce_1965}, 
when using the new planetary radar at Arecibo Observatory in Puerto Rico.
Subsequent observations confirmed that contrary to previous expectations
\citep{Schiaparelli_1890,Defrancesco_1988}, the
rotation of this planet was not synchronous with the orbital mean motion,
but presented a peculiar 3/2 resonant equilibrium
\citep{McGovern_etal_1965,Colombo_1965}. 
Within a year of the discovery the stability of this equilibrium became
understood, as the result of the solar torque on Mercury's quadrupolar
moment of inertia combined with an eccentric orbit
\citep{Colombo_Shapiro_1966,Goldreich_Peale_1966,Counselman_Shapiro_1970}.
However, the reason why this state initially arose remained unsatisfactory for a
long time.

Mercury like all the other planets in our Solar System is supposed to have had
an initially rapid spin, that was slowed down by the continuous action of the
intense solar tides \citep{Darwin_1880,Peale_1974,Peale_1976,Burns_1976}. 
As the spin rate approaches the orbital mean motion, it will cross a series of
resonances. 
In their work, \citet{Goldreich_Peale_1966} have shown that,
since the tidal strength depends on the planet's rotation rate, it creates an
asymmetry in the tidal potential that allows the capture into these 
spin-orbit resonances. They also computed the capture probability 
into these resonances for a single crossing, 
and found that for the present eccentricity value of Mercury ($ e = 0.206 $),
and  unless one uses an unrealistic tidal model with constant torques, 
the probability of capture into the present 3/2 spin-orbit resonance is on the
low side, at most about 7\%, which  remained somewhat unsatisfactory. 

Later, \citet{Correia_Laskar_2004} have shown that, 
as the orbital eccentricity of Mercury is chaotically 
varying, with some excursions to high values, the rotation 
rate of the planet can be accelerated again, and 
the 3/2 resonance could have been crossed many times in the past. 
Performing a statistical study of the past evolution of Mercury's orbit,
over 1000 cases, it was demonstrated that
capture into the 3/2 spin-orbit resonant
state is in fact, and without the need 
of a specific core-mantle effect, the most probable final outcome of the planet's evolution,
occurring about 55.4\% of the time.

\citet{Goldreich_Peale_1967} had nevertheless pointed out that the
probability of capture could be greatly enhanced if a planet has a molten core.
In 1974, the discovery of an intrinsic magnetic field by the Mariner 10
spacecraft  seemed to imply the existence of a conducting molten core
\citep{Ness_etal_1974,Ness_etal_1975}, and more recently \citet{Margot_etal_2007}
confirmed its existence by using radar observations.
Core-mantle friction is then an effect to take into account and we expect an
increment in the capture probabilities for the 3/2 resonance.
However, according to \citet{Goldreich_Peale_1967}, 
this also increases the capture probability  in all the previous resonances.
\citet{Peale_Boss_1977} indeed remarked  that 
only very specific values of the core
viscosity allow to avoid the 2/1 resonance and permit capture in the 3/2.

More recently \citep{Correia_Laskar_2009}, it was shown that, as the chaotic evolution
of Mercury's orbit can also drive its eccentricity to very low values during the
planet's history, any previous capture can be destabilized whenever the
eccentricity becomes lower than a critical value, except for the 1/1 resonance. 
Including the core-mantle friction effect combined with the chaotic evolution of
the eccentricity, it was found that
the spin ends 99.8\% of the time captured in a spin-orbit resonance, 
mainly distributed by the following three configurations: 5/2
(22\%), 2/1 (32\%) and 3/2 (26\%).
Although in this case the present 3/2 spin-orbit resonance is not the most probable outcome, 
it was also shown that the probability of ending up in this resonance can be increased up
to 55\% or 73\%, if the eccentricity of Mercury in the past has descended
below the critical values 0.025 or 0.005, respectively.

The present paper is the continuation of the previous
works \citep{Correia_Laskar_2004, Correia_Laskar_2009}.
Here, we revisit in full detail the theory of dissipative effects (tides and
core-mantle friction) and mechanisms for capture in resonance, in order to
compare with the results from previous studies, in particular those from
\citet{Goldreich_Peale_1966,Goldreich_Peale_1967}. 
We are particularly interested in considering the effect of a non-zero
obliquity, since all previous studies assumed that the obliquity was close to
zero, because this corresponds to the final evolution from dissipative effects.
However, at the time of the first resonance crossing, the  obliquity may still
be significant, which can lead to important modifications in capture probabilities.
In a forthcoming paper we will present the full dynamics of the spin with
planetary perturbations, which are not included in the present work.

In the next section, we give the averaged conservative equations in a suitable
form for simulations of the long-term variations of Mercury's spin, including
the resonant terms and the precession motion. In Section 3 we define a model for
taking into account the tidal effects including the obliquity contribution.
Section 4 is devoted to the analysis of  the core-mantle friction.
This effect can be divided in two parts, one resulting from the
libration around the resonance (included by \citet{Goldreich_Peale_1967}), and a
non-resonant term which depends on the obliquity.
While the first term tends to increase the chances of capture, the other has the
opposite effect.
In Section 5 we revisit the theory of spin-orbit resonances and Section 6 is
devoted to dynamical equation analysis and its implications. 
In Section 7 we perform some numerical integrations, tracking the spin evolution
from its origin to the present and computing the chances of capture in different
resonances.

\section{Conservative motion}

We will first omit the dissipative effects, and describe the spin motion of the
planet in a conservative framework, including the obliquity contributions.
The motion equations will be easily obtained from the Hamiltonian function
of the total rotational energy of the planet.

Mercury is considered here as an homogeneous rigid body with mass $ m $ and
moments of inertia $ A \le B < C $, supported by the reference frame $ ( \vv{\ii},
\vv{\jj}, \vv{\kk} ) $, fixed with respect to the planet's figure.
We do the gyroscopic approximation, i.e., we merge the axis of principal inertia and the axis of rotation, since for a
long-term study we are not interested in nutations.
Let $ \vv{L} $ be the total rotational angular momentum and $ ( \vv{\mathrm{I}},
\vv{\mathrm{J}},\vv{\KK} ) $ a  reference frame linked to the orbital plane
(where $ \vv{\KK} $ is the normal to this plane). 
The angle between $ \vv{\kk} $ and $ \vv{\KK} $ is the obliquity, $ \ve $, and
thus, $ \cos \ve = \vv{\kk} \cdot \vv{\KK} $.
The Hamiltonian of the motion can be written
using canonical Andoyer's action variables ($ L , X $) and their conjugate
angles ($ \theta , - \psi $) \citep{Andoyer_1923,Kinoshita_1977}.
$ L = \vv{L} \cdot \vv{\kk} = C \omega $ is the projection of the angular
momentum on the $ C $ axis, with rotation rate $ \omega = \dot \theta - \dot \psi
\cos \ve $ and $ X = \vv{L}
\cdot \vv{\KK} $ its projection on the normal to the ecliptic; 
$ \theta $ is the hour angle between the equinox of date and a fixed point of
the equator, and $ \psi $ is the general precession angle (Fig.\ref{Fig01}).
\prep{\FigA}

\subsection{Gravitational potential}

\llabel{030214c}

The gravitational potential $ \cV $ generated by the planet at a generic point
of the space $ \vv{r} $  is given by \citep[e.g.][]{Tisserand_1891,Smart_1953}:  
\begin{eqnarray}
\cV (\vv{r}) = - \Frac{G m}{r} & + & \Frac{G
(B - A)}{r^3} P_2 ( \vv{\hat r} \cdot \vv{\jj} ) \crm & + &  \Frac{G (C - A)}{r^3}
P_2 ( \vv{\hat r} \cdot \vv{\kk} ) \ , \llabel{090418a}
\end{eqnarray}
where $ G $ is the gravitational constant, $ \vv{\hat r} = \vv{r} / r $ and $
P_2 (x) = (3 x^2 -1)/2 $ are the Legendre polynomials of degree two.
The potential energy $ \cU $ when orbiting a central star of mass $ m_\odot $ 
is then:
\be
\cU = m_\odot \cV (\vv{r}_{}) \llabel{021209a} \ .
\ee

For a planet evolving in a
non-perturbed keplerian orbit, we write:
\be
\vv{\hat r} = \cos (\varpi + v) \vv{\mathrm{I}} + \sin (\varpi + v)
\vv{\mathrm{J}} \ , \llabel{090418b}
\ee
where $ \varpi $ is the longitude of the perihelion and $ v $ the true anomaly.
Thus, transforming the body equatorial frame $ ( \vv{\ii}, \vv{\jj}, \vv{\kk} ) $ in the
ecliptic one $ ( \vv{\mathrm{I}}, \vv{\mathrm{J}}, \vv{\KK} ) $, we obtain
\citep[e.g.][]{Correia_2006}: 
\be
 \left\{ 
  \begin{array}{l l}
   \vv{\hat r} \cdot \vv{\jj}  = - \cos w \sin \theta + \sin w \cos \theta \cos \ve 
     \ , \crm
   \vv{\hat r} \cdot \vv{\kk}  = - \sin w \sin \ve \ ,
  \end{array} 
 \right. \llabel{090418c}
\ee 
where $ w = \varpi + \psi + v $ is the true longitude of date.
The expression for the potential energy (\ref{021209a}) becomes:
\begin{eqnarray} 
\cU & = & - \Frac{G m m_\odot}{r_{}} + \Frac{G C m_\odot}{r_{}^3} E_d P_2 ( \sin
w \sin \ve ) \crm & & - \Frac{3 G m_\odot}{8 r_{}^3} (B - A) \, F (\theta, w, \ve)
\ , \llabel{030123b} 
\end{eqnarray}  
where
\begin{eqnarray} 
& F (\theta, w, \ve) =  2 \cos ( 2 \theta - 2 w ) \cos^4 \left( \Frac{\ve}{2}
\right) \quad \quad \quad \quad \quad  &  \crm 
& + 2 \cos ( 2 \theta + 2 w ) \sin^4 \left( \Frac{\ve}{2} \right) + \cos (2
\theta) \sin^2 \ve & \llabel{030123c}
\end{eqnarray}  
and
\be 
E_d = \frac{C - \frac{1}{2} (A + B)}{C} = \frac{ k_f R^5 }{3 G C}
\omega^2 + \delta  E_d \ . \llabel{050109a}
\ee 
$ R $ is the planet's radius and  $ k_f $ the fluid Love number (pertaining to a
perfectly fluid body with the same mass distribution as the actual planet). 
$ E_d $ is the dynamical ellipticity, the first part of this expression
corresponding to the flattening in hydrostatic equilibrium \citep{Lambeck_1980},
and $ \delta E_d $ to the departure from this equilibrium. 

\subsection{Averaged potential} 

\llabel{030123a}

Since we are only interested in the study of the long-term motion, we 
will average the potential energy $\cU$ over the rotation angle $\theta$ and the
mean anomaly $ M $, after expanding the true anomaly $ v $ in series of the
eccentricity $e$ and mean anomaly. 
However, 
when the rotation frequency $ \omega \simeq \dot \theta $ and the mean motion $ n = \dot M $
are close to resonance ($ \omega \simeq p n $, for a semi-integer\footnote{We have retained the use
of semi-integers for better  comparison with previous results} value $p$),
we must retain the terms with argument $ 2(\theta - p M) $ in the expansions 
\be 
\frac{\cos ( 2 \theta )}{r_{}^3} = \frac{1}{a^3} \sum^{+\infty}_{p=-\infty} G
( p , e ) \cos (2 \theta - 2 p M) \llabel{061120ga}
\ee
and  
\be 
\frac{\cos ( 2 \theta - 2 v )}{r_{}^3} = \frac{1}{a^3} \sum^{+\infty}_{p=-\infty} H
( p , e ) \cos (2 \theta - 2 p M) \ , \llabel{061120gb}
\ee  
where  $ a $ is the semi-major axis of the planet's orbit and 
the functions $ G (p, e) $ and $ H (p, e) $ are power
series in $ e $ (Tab.~\ref{TAB1}).
\prep{\TabA}
The averaged  potential $ \overline \cU $ becomes:
\begin{eqnarray}
\Frac{\overline \cU}{C} & = & 
- \alpha \Frac{\omega \, x^2}{2} - \Frac{\beta}{4} \left[ (1-x^2) \, G (p,e) \cos 2 (\theta
- \mmu M) \phantom{\Frac{.}{.}} \right. \crm
&  & + \Frac{(1+x)^2}{2} \, H (p,e) \cos 2 (\theta - \mmu M - \phi) \crm
&  & + \left. \Frac{(1-x)^2}{2} \, H (-p,e) \cos 2 (\theta - \mmu M + \phi)
\right] \ , \llabel{030804a} 
\end{eqnarray}
where $ x = X / L = \cos \ve $, $ \phi = \varpi + \psi $,
\be 
\alpha = \Frac{3 G m_\odot}{2 a^3 (1-e^2)^{3/2}} \Frac{E_d}{\omega} \simeq
\Frac{3}{2} \Frac{n^2}{\omega} (1-e^2)^{-3/2} E_d \llabel{061120a}
\ee
is the `precession constant' and
\be 
\beta = \Frac{3 G m_\odot}{2 a^3} \Frac{B - A}{C} \simeq \Frac{3}{2}
n^2 \Frac{B - A}{C} \ . \llabel{030123g}
\ee
We can rewrite expression (\ref{030804a}) simplified as:
\be
\Frac{\overline \cU}{C} = - \alpha \Frac{\omega \, x^2}{2} - \Frac{\betal}{2} 
\cos 2 (\theta - \mmu M - \phil ) \ , \llabel{030123e}
\ee
where the amplitude $ \betal $ and the phase angle $ \phil $ are functions
depending on both $ x $ and $ \psi $, whose expressions are given in appendix~\ref{ApenB}.

\subsection{Equations of motion}

\llabel{040811x}

The Andoyer  variables ($ L $, $ \theta$) and ($ X $, $ - \psi $) are 
canonically conjugated and thus
\be 
\frac{d L}{d t} = -\frac{\partial \overline \cU}{\partial \theta} \ ;
\quad \frac{d X}{d t} = \frac{\partial \overline \cU}{\partial \psi} \ ;
\quad \frac{d \psi}{d t} = - \frac{\partial \overline \cU}{\partial X} \ .
\llabel{021014c}
\ee
Despite their practical use, Andoyer's variables do not give a clear view
of the obliquity variations. Since $ \cos \ve = X / L $ they can be obtained as:
\begin{eqnarray} 
\sin \ve \, \frac{d \ve}{d t} = \frac{1}{L} \left(
\frac{X}{L} \frac{d L}{d t} - \frac{d X}{d t} \right) = - \frac{1}{L}
\left[ x \frac{\partial \overline \cU}{\partial \theta} 
+ \frac{\partial \overline \cU}{\partial \psi} \right] \llabel{050602a} \ .
\end{eqnarray}
Then, from equation (\ref{030804a}) we get:
\be
\frac{d L}{d t} = - C \betal \sin 2 (\theta - p M - \phil) \ ,
\llabel{021014e}
\ee
\be
\frac{d \ve}{d t} = - \alpha_r \sin \ve \cos 2 (\theta - p M -
\phi_r) \ , \llabel{021016f}
\ee
and
\be
\frac{d \psi}{d t} = \alpha \, x + \alpha_r \sin 2
(\theta - p M - \phi_r)  \ , \llabel{021016d}
\ee
where $ \alpha_r $ and $ \phi_r $ are functions depending on both $ x $ and $
\psi $, whose expressions are given in appendix~\ref{ApenB}.
For non-resonant motion the previous equations simplify as:
\be 
\frac{d L}{d t} = \frac{d \ve}{d t} = 0 \quad \mathrm{and}
\quad \frac{d \psi}{d t} = \alpha \cos \ve \ . \llabel{050214b}
\ee
The planet spin motion reduces to the precession of the spin vector about the
normal to the orbital plane with rate $ \alpha \cos \ve $.
In general we have $ \alpha \gg \alpha_r $ and the precession rate including the
resonant motion (Eq.\ref{021016d}) is nearly the same as the non-resonant case
(Eq.\ref{050214b}).

\section{Tidal effects}

\llabel{030214b}

Tidal effects arise from differential and inelastic deformations of the planet
due to the gravitational effect of a perturbing body. 
Their contributions to the spin variations are based on a very
general formulation of the tidal potential, initiated by George H. Darwin
(1880). The attraction of a body with mass $ m_{\odot} $
at a distance $ r_{} $ from the center of mass of the planet can be expressed
as the gradient of a scalar potential $\cV'$, which is a sum of Legendre polynomials:
\be 
\cV' = \sum_{l=2}^{\infty} \cV_l' = - \frac{G m_{\odot}}{r_{}}
\sum_{l=2}^{\infty} \left(  \frac{r'}{r_{}} \right)^l P_l (\cos S) \ ,
\llabel{090419b}
\ee 
where $ r' $ is the radial distance from the planet's center, and $ S $ the
angle between $\vv{r}$ and $\vv{r}'$. 
The distortion of the planet by this potential gives rise to a tidal potential,
\be 
\cV^g = \sum_{l=2}^{\infty} \cV_l^g \ , \llabel{090419c}
\ee
where $ \cV_l^g = k_l \cV_l' $ at the planet's surface and $ k_l $
is the Love number for potential.  
Since the tidal potential $ \cV_l^g $ is an $ l $th degree harmonic, exterior to
the planet it must be proportional to $ r^{-l-1} $ (solution of a Dirichlet 
problem).
Furthermore, as upon the surface $ r' = R \ll r_{} $, we can retain in
the expansion only its first term, $ l = 2 $:
\be 
\cV^g = - k_2 \frac{G m_{\odot}}{R} \left( \frac{R}{r_{}} \right)^3
\left( \frac{R}{r'} \right)^3 P_2 (\cos S) \ . \llabel{021010a}
\ee 

In general, imperfect elasticity will cause the phase angle of $ \cV^g $ to lag
behind that of $ \cV' $ \citep{Kaula_1964} by an angle $ \delta (\sigma) $ such
that:  
\be 
\delta (\sigma) = \frac{\sigma \Delta t (\sigma)}{2} \ , \llabel{090419d}
\ee
$ \Delta t (\sigma) $ being the time lag associated to the tidal frequency $
\sigma $ (a linear combination of the inertial rotation rate $ \omega
$ and the mean orbital motion $ n $). 

\subsection{Equations of motion}

\llabel{021205a}

Expressing the tidal potential given by expression (\ref{021010a}) in
terms of Andoyer angles $( \theta, \psi )$, we then easily obtain its contribution
to the spin evolution as:
\be
   \frac{d L}{d t} =  - m' \frac{\partial \cV^g}{\partial \theta} 
   \ ; \quad
   \frac{d X}{d t} =  m' \frac{\partial \cV^g}{\partial \psi} 
   \ ; \llabel{021010b}
\ee
where $ m' $ is the mass of the interacting body.
As we are interested here in the study of the secular evolution of the spin,
we will average (\ref{021010b}) over the periods of mean anomaly, longitude of
node and perihelion of the perturbing body. 
When the interacting body is the same as the perturbing one $( m' = m_\odot )$, we
obtain:
\be
   \Frac{d L}{d t} = - \Frac{G m_{\odot}^2 R^5}{a^6} \sum_{\sigma}
                     b (\sigma) \cQ^L_\sigma (x, e) \ ,
 \llabel{021010c1} 
\ee
\be                   
   \Frac{d \ve}{d t} = - \Frac{G m_{\odot}^2 R^5}{a^6}
          \Frac{\sin \ve}{L} \sum_{\sigma} b (\sigma) \cQ^\ve_\sigma (x, e) \ ,
 \llabel{021010c2} 
\ee
where the coefficients $ \cQ_\sigma (x, e) $ are polynomials in the
eccentricity \citep{Kaula_1964}. 
The factors $ b (\sigma) $ are related to the dissipation of the mechanical energy
of tides in the planet's interior responsible for the time delay $ \Delta t
(\sigma) $ between the position of ``maximal tide'' and the sub-solar point. 
They are related to the phase lag $ \delta (\sigma) $ as:
\be  
b (\sigma) = k_2 \sin 2 \delta (\sigma) = k_2 \sin
\left( \sigma \Delta t (\sigma) \right) \ . \llabel{021010e}
\ee 
Dissipation equations (\ref{021010c1}) and (\ref{021010c2}) must be invariant under the change $
(\omega, \ve) $ by $ (- \omega, \pi - \ve) $ which imposes that $ b (\sigma) = -
b (- \sigma) $, that is, $ b (\sigma) $ is an odd function of $\sigma$.
Although mathematically equivalent, the couples $ (\omega, \ve) $ and $ (-
\omega, \pi - \ve ) $ correspond to two different physical situations
\citep{Correia_Laskar_2001}.

\subsection{Dissipation models}

\llabel{021024j}

The dissipation of the mechanical energy of tides in the planet's interior is
responsible for the phase lags $ \delta (\sigma) $.
A commonly used dimensionless measure of tidal damping is the quality factor $
Q $ \citep{Munk_MacDonald_1960}, defined as the inverse of the ``specific''
dissipation and related to the phase lags by
\be 
Q (\sigma) = \frac{2 \pi E}{\Delta E} = \cot 2 \delta (\sigma) \ ,
\llabel{090419e}
\ee
where $ E $ is the total tidal energy stored in the planet, and $ \Delta
E $ the energy dissipated per cycle. We can rewrite (\ref{021010e}) as:
\be 
b (\sigma) = \frac{ k_2 \, \mathrm{sign}(\sigma)}{\sqrt{Q^2 (\sigma) + 1}}
\simeq \mathrm{sign}(\sigma) \frac{k_2}{Q (\sigma)} \ .
\ee 
The present $ Q $ value of the planets in the Solar system
can be estimated from orbital measurements,
but as rheology of the planets is  badly known,
the dependence of $b(\sigma)$ on the tidal frequency $ \sigma $
is subject to various approximations.

\subsubsection{The visco-elastic model}

\llabel{021120y}

\citet{Darwin_1908} assumed that the planet behaves like a Maxwell
solid\footnote{A material is called Maxwell solid when it responds to stresses
like a massless, damped harmonic oscillator. It is characterized by a rigidity
(or shear modulus) $ \mu_e $ and by a viscosity $ \upsilon_e $. A Maxwell solid
behaves like an elastic solid over short time scales, but flows like a fluid
over long periods of time. This behavior is also known as elasticoviscosity.} of
constant density $ \rho $, and found:
\be
b (\sigma) = k_f \frac{\tau_b - \tau_a}{1 + (\tau_b \, \sigma)^2} \sigma \ ,
\llabel{090419f}
\ee
where $ \tau_a = \upsilon_e / \mu_e $ and $ \tau_b $ are the time constants for
damping of the body tides,
\be
\tau_b = \tau_a ( 1 + 19 \mu_e R / 2 G m \rho ) \ . \llabel{090419g}
\ee

The visco-elastic model is a very realistic approximation of the planet's
deformation with the tidal frequency.
However, when replacing expression (\ref{090419f}) into the dynamical
equations (\ref{021010c1}) and (\ref{021010c2}) we get an infinite sum of terms.
This problem can be solved by using simplified versions of the visco-elastic model
for specific values of the tidal frequency $ \sigma $.
For instance, when $\sigma$ is small, $ (\tau_b \, \sigma)^2$ can be neglected in
(\ref{090419f}) and $b(\sigma)$ becomes proportional to $\sigma$.

\subsubsection{The viscous or linear model}

\llabel{021120a}

In this model, 
it is assumed that the response time delay to the perturbation is 
independent of the tidal frequency, i.e., the position of the ``maximal
tide'' is shifted from the sub-solar point by a constant time lag $ \Delta t $
 \citep{Mignard_1979,Mignard_1980}. As usually we have  $ \sigma \Delta t \ll 1 $,
this model becomes linear:
\be 
b^g (\sigma) = k_2 \sin (\sigma \Delta t) \simeq k_2 \, \sigma \Delta t \ .
\ee  
Substituting the above formula into expressions (\ref{021010c1}) and (\ref{021010c2}), we simplify
the motion equations as (appendix~\ref{ApenC}): 
\be
 \left\{ 
  \begin{array}{l l}
   \Frac{d L}{d t} = - C K
   \left[ \Frac{1 + x^2}{2} \, \Omega (e) \Frac{\omega}{n} - x N (e) \right] \ ,
                    \crm 
   \Frac{d \ve}{d t} = K \Frac{\sin \ve}{\omega} \left[ x \, \Omega (e)
      \Frac{\omega}{2 n} - N(e) \right]  \ ,
  \end{array} 
 \right.
 \llabel{040811a} 
\ee
where
\be
\Omega (e) = \Frac{1 + 3 e^2 + 3 e^4 / 8}{(1 - e^2)^{9/2}} \ , \llabel{030218a}
\ee
\be
N (e) = \Frac{1 + 15 e^2 / 2 + 45 e^4 / 8+ 5 e^6 / 16}{(1 - e^2)^{6}} \ ,
\llabel{030218b}
\ee
\be
K = n^2 \, \frac{3 \, k_2}{\xi \, Q} \left(\frac{m_\odot}{m}\right) 
\left(\frac{R}{a}\right)^3 \ , \llabel{eq3}
\ee
$ Q^{-1} = n \Delta t $ and $ \xi = C / (m R^2) $.
The viscous model is a particular case of the visco-elastic model and is
specially adapted to describe the behavior of planets in slow rotating regimes
($ \omega \sim n $).

\subsubsection{The constant-$Q$ model}

\llabel{021024i}

Since for the Earth, $ Q $   changes by less than an order of magnitude between the
Chandler wobble period (about 440 days) and seismic periods of a few
seconds, it is also common to treat the specific dissipation as independent of
frequency. Thus,
\be 
b^g (\sigma) \simeq \mathrm{sign}(\sigma) k_2 / Q \ . \llabel{090407a}
\ee
For long-term evolutions and slow rotating planets, this model is not
appropriate as it gives rise to discontinuities for $ \sigma = 0 $.
However, it can be used for periods of time where the tidal frequency does
not change much, as is the case for fast rotating planets.  
Substituting Eq.(\ref{090407a}) into expressions (\ref{021010c1}) and
(\ref{021010c2}), we can simplify the motion equations as:
\be
 \left\{ 
  \begin{array}{l l}
   \Frac{d L}{d t} = - s_g \Frac{C K}{16} \left( 5 + 6 x^2 - 3 x^4
   \right) \Omega (e) \ , \crm 
   \Frac{d \ve}{d t} = \Frac{K}{16} \, \Frac{\sin
        \ve}{\omega} \left( 3 - 7 s_g \, x  -3 x^2 + 3 s_g \, x^3 \right) \Omega (e) \ ,
  \end{array} 
 \right.
 \llabel{040811r} 
\ee
where $ s_g = \mathrm{sign} (L) = \mathrm{sign} (\omega) $.

\section{Core-mantle friction effect}

\llabel{sectioCoreMantle}

The Mariner 10 flyby of Mercury revealed the presence of an intrinsic magnetic
field, which is most likely due to motions in a conducting fluid inner core
\citep[for a review see][]{Ness_1978,Spohn_etal_2001}.
Subsequent observations made with Earth-based radar provided strong evidence
that the mantle of Mercury is decoupled from a core that is at least
partially molten \citep{Margot_etal_2007}.
If there is slippage between the liquid core and the mantle, a second source of
dissipation of rotational energy results from friction occurring at the
core-mantle boundary. 
Indeed, because of their different shapes and densities, the core and the mantle
do not have the same dynamical ellipticity and the two parts tend to precess
at different rates \citep{Poincare_1910}.
This tendency is more or less counteracted by different interactions produced at
their interface: the torque $ \vv{N} $ of non-radial inertial pressure forces of
the mantle over the core provoked by the non-spherical shape of the interface;
the torque of the viscous (or turbulent) friction between the core and the
mantle; the torque of the electromagnetic friction, caused by the interaction
between electrical currents of the core and the bottom of the magnetized mantle.

\subsection{Equations of motion}

We will adopt henceforward a model for the planet which is an extension of the
model from \citet{Poincare_1910} of a perfect incompressible and homogeneous
liquid core with moments of inertia $ A_c = B_c < C_c $ inside an homogeneous
rigid body with moments of inertia $ A_m \le B_m < C_m $, supported by the same
reference frame $ ( \vv{\ii}, \vv{\jj}, \vv{\kk} ) $, fixed with respect to the
planet's figure (Fig.\ref{Fig01}).
The combined effects of inertial and frictional coupling across the ellipsoidal
core-mantle boundary are taken into account, assuming laminar flow.

Denoting $ \vv{\delta} = \vv{\omega} - \vv{\omega}_c $ the differential
rotation between the core and the mantle, we can write the non-radial inertial
pressure torque of the mantle on the core in a general formulation to first
order in the core dynamical ellipticity, $ E_c $, as \citep{Rochester_1976,Sasao_etal_1980,Pais_etal_1999}:
\be 
\vv{N} = \vv{\omega}_c \times \vv{L}_c = C_c E_c \, \omega \,
\vv{\kk} \times \vv{\delta} \ , \llabel{050514a}
\ee
where $ \vv{L}_c = \mm{I}_c \cdot \vv{\omega}_c $ is the core angular momentum
with $ \mm{I}_c = \mathrm{diag} (A_c,A_c,C_c) $ its tensor of inertia. 

The two types of friction torques (viscous and electromagnetic) depend
on the differential rotation between the core and the mantle and can be
expressed by a single effective friction torque, $ \vv{\Gamma} $.
As a general expression for this torque we adopt
\citep{Rochester_1976,Sasao_etal_1980,Mathews_Guo_2005}
\be 
\vv{\Gamma} = C_c \, ( \kappa + \kappa' \, \vv{\kk} \times ) \,
\vv{\delta}  \ , \llabel{020410a} 
\ee  
where $ \kappa $ and $ \kappa' $ are effective coupling parameters.
They result either from viscous and electromagnetic stresses at
the core-mantle interface and can be written as a sum of these two effects: $
\kappa = \kappa_\mathrm{vis} + \kappa_\mathrm{em} $ and $
\kappa' = \kappa_\mathrm{vis}' + \kappa_\mathrm{em}' $.
Recent estimations of these coefficients can be found in the works of
\citet{Mathews_Guo_2005} and \citet{Deleplace_Cardin_2006}.
In the simplified case of no magnetic field, the coupling parameters are only
given by the viscous friction contributions, which can be simplified as
\citep{Noir_etal_2003,Mathews_Guo_2005}: 
\be 
\kappa_\mathrm{vis} = 2.62 \sqrt{\nu \md \omega \md } / R_c  \quad \mathrm{and} \quad  
\kappa_\mathrm{vis}' = 0.259 \sqrt{\nu \md \omega \md } / R_c \ , \llabel{V14a} 
\ee
where $ R_c $ is the core radius and $ \nu $ the kinematic viscosity, which
is poorly known. Even in the case of the Earth, the uncertainty in $ \nu $
covers about 13 orders of magnitude \citep{Lumb_Aldridge_1991}, the best
estimate so far being $ \nu \simeq 10^{-6} \, \mathrm{m}^2 \mathrm{s}^{-1} $
\citep{Gans_1972,Poirier_1988,deWijs_etal_1998}.

Since the derivative of the angular momentum is given by the sum of external
torques, the contribution of the core-mantle friction is the solution of
the system: 
\be
 \left\{ 
  \begin{array}{l l}
   \Frac{d \vv{L}_m }{d t} = \vv{P} + \vv{T} - \vv{N} - \vv{\Gamma}  \ , \crm
   \Frac{d \vv{L}_c }{d t} = \vv{N} + \vv{\Gamma}  \ , 
  \end{array} 
 \right.
\llabel{V12} 
\ee
where $ \vv{P} = \vv{r} \times \nabla {\cal U} $ is the precession torque and
$ \vv{T} = \vv{r} \times m' \nabla {\cal V}^g $ the tidal torque.
$ \vv{L}_m $ denotes the mantle's angular momentum,
\be
\vv{L}_m = C_m \, \vv{\omega} = C_m \, \omega \, \vv{\kk}  \llabel{060831a}
\ee
and the total angular momentum variations are given by:
\be
\Frac{d \vv{L}}{d t} = \Frac{d \vv{L}_m}{d t} + \Frac{d \vv{L}_c}{d t} = \vv{P}
+ \vv{T} \ . \llabel{060113a} 
\ee
$ \vv{L} $ precesses around $ \vv{\KK} $, a normal vector to the orbital
plane, with angular velocity $ \vv{\Omega} $ given by:
\be
\vv{\Omega} = - \dot \psi \vv{\KK} + \dot \ve \, \vv{p} \ , \llabel{061115a}
\ee
where $ \dot \ve $ accounts for the secular effects resulting from tides and
core-mantle friction and $ \vv{p} \equiv \vv{\KK} \times \vv{\kk} \, / \sin \ve $ is
the unit vector along the direction of the averaged precession. 
Thus,
\be
\Frac{d \vv{L}}{d t} = \vv{\Omega} \times \vv{L} = \vv{\Omega} \times \left(
C \omega \vv{\kk} - \mm{I}_c \cdot \vv{\delta} \right) \ . \llabel{061115b} 
\ee
Projecting it over 
\be 
\vv{q} \equiv \vv{\kk} \times \vv{p} = \frac{\vv{\KK} - \vv{\kk} \cos \ve}{\sin \ve} \
, \llabel{061115c}
\ee
we get
\begin{eqnarray}
 ( \vv{\Omega} \times \vv{L} ) \! \! \! & \cdot & \! \! \! ( \vv{\kk} \times
 \vv{p}) \ \ = \ \ 
(\vv{\Omega} \cdot \vv{\kk} )( \vv{L} \cdot \vv{p} ) - (\vv{\Omega} \cdot
\vv{p} )( \vv{L} \cdot \vv{\kk} )  \crm & = &  - \dot \psi \cos \ve \, (-
\vv{p} \cdot \mm{I}_c \cdot \vv{\delta} ) - \dot \ve \, ( C \omega - \vv{\kk}
\cdot \mm{I}_c \cdot \vv{\delta} ) \crm & = & \dot \psi \cos \ve A_c \delta_p -
\dot \ve ( C \omega - C_c \delta_k ) \ , \llabel{061115d}
\end{eqnarray}
with $ \delta_\kk = \vv{\delta} \cdot \vv{\kk} $ and $ \delta_p = \vv{\delta}
\cdot \vv{p} $.
Then, assuming $ \delta_k \ll \omega $ we have from expression (\ref{060113a}) 
\be
\dot \ve = - \Frac{P_\qq}{C \omega} - \Frac{T_\qq}{C \omega} + \Frac{\dot \psi
\cos \ve A_c \delta_p}{C \omega} \ , \llabel{061115e}
\ee
where $ - P_\qq / C \omega $ and $ - T_\qq / C \omega $ are respectively given
by expressions (\ref{021016f}) and (\ref{021010c2}).
The value of $ \delta_p $ in the case of uniform precession is
\citep{Rochester_1976,Pais_etal_1999,Correia_2006}:
\be
\delta_p = \Frac{- \kappa \, \omega \, \dot \psi \sin \ve}{\kappa^2 + (\kappa' +
E_c \omega)^2} \ . \llabel{061115f} 
\ee
In the case of a non-uniform precession, as it can be the case of Mercury close
to a spin-orbit resonance, \citet{Correia_2006} shows that the asymmetric terms
in $(B-A)$ are periodic and average to zero, and we can still use the
previous expression for $ \delta_p $.
Using $ c_c = C_c / C $, $ A_c \simeq C_c $ and $ \dot \psi \simeq \alpha \cos
\ve $, we finally have for the obliquity variations with core-mantle friction:
\be
\dot \ve = - \Frac{P_\qq}{C \omega} - \Frac{T_\qq}{C \omega} - K_f \cos^3 \ve
\sin \ve \ , \llabel{061122b}
\ee
with
\be
K_f = \Frac{c_c  \kappa \, \alpha^2 }{\kappa^2 + (\kappa' + E_c \omega)^2}
\ , \llabel{061122c} 
\ee
which is always a positive quantity.


The rotation rate variations of the mantle and the core can be
obtained by projecting both equations (\ref{V12}) onto $ \vv{\kk} $:
\be
\Frac{d}{d t} ( \vv{L}_i \cdot \vv{\kk} ) = \Frac{d \vv{L}_i}{d t} \cdot
\vv{\kk} + \vv{L}_i \cdot \Frac{d \vv{\kk}}{d t} = \Frac{d \vv{L}_i}{d t} \cdot
\vv{\kk} + \vv{L}_i \cdot ( \vv{\Omega} \times \vv{\kk} ) \llabel{061115g} \ .
\ee
For the mantle, $ \vv{L}_i $ is given by expression (\ref{060831a}) and then
\be
C_m \Frac{d \omega}{d t} = P_\kk + T_\kk - C_c \kappa \, \delta_\kk \ ,
\llabel{050428b}  
\ee
where $ P_\kk $ and $ T_\kk $ are respectively given by
expressions (\ref{021014e}) and (\ref{021010c1}).
For the core $ \vv{L}_i = \vv{L}_c = \mm{I}_c \cdot \vv{\omega}_c $ and we have
\begin{eqnarray}
C_c \Frac{d \omega_c^\kk}{d t} & = & C_c \kappa \, \delta_\kk
+ \vv{L}_c \cdot (-\dot \psi \sin \ve \, \vv{p} - \dot \ve \, \vv{q} ) \crm
& = & C_c \kappa \, \delta_\kk + A_c \dot \psi \sin \ve \, \delta_p + A_c \dot
\ve \, \delta_\qq \ ,
\llabel{061114a} 
\end{eqnarray}
where $ \omega_c^\kk = \vv{\omega}_c \cdot \vv{\kk} = \omega - \delta_\kk $
and $ \delta_p $ is given by expression (\ref{061115f}).
We can also neglect the term in $ A_c \, \dot \ve \, \delta_\qq $ because according
to expression (\ref{061115e}) its average is a second order term in $ \delta
$. Thus, using $ A_c \simeq C_c $ we have
\be
\Frac{d \omega_c^\kk}{d t} = \kappa \, \delta_\kk - \Frac{\omega K_f}{c_c} \cos^2 \ve \sin^2 \ve \ .
\llabel{061116z} 
\ee

\subsection{Differential rotation}

\llabel{061129c}

Combining equations (\ref{050428b}) and (\ref{061114a}) we find a
differential equation for $ \delta_\kk $,
\be
\Frac{d \delta_\kk}{d t} = - \kappa_m \, \delta_\kk + \Frac{P_\kk}{C_m} +
\Frac{T_\kk}{C_m} - \Frac{A_c}{C_c} \dot \psi \sin \ve \, \delta_p \ ,
\llabel{061115z} 
\ee
where $ \kappa_m = \kappa \, C / C_m $. Its solution allows us completely to 
determine the spin of the mantle (Eq.\ref{050428b}) without needing the core
variations (Eq.\ref{061114a}): 
\be
\delta_\kk (t) = \ei^{-\kappa_m t} \int \left( \Frac{P_\kk}{C_m} +
\Frac{T_\kk}{C_m} - \Frac{A_c}{C_c} \dot \psi \sin \ve \, \delta_p \right)
\ei^{\kappa_m t} d t \ . \llabel{061116a} 
\ee
The secular variations to the spin can be seen as constant for short-periods
of time.
Thus, since among all the contributions inside the integral of previous
expression only $ P_\kk $ is not secular, we have
\be
\delta_\kk (t) = \Frac{1}{\kappa} \left( \Frac{T_\kk}{C} - \Frac{C_m}{C} \dot
\psi \sin \ve \, \delta_p \right) + P (t) \ , \llabel{061116b} 
\ee
with
\be
P (t) = \ei^{-\kappa_m t} \int \Frac{P_\kk}{C_m} \, \ei^{\kappa_m t} d
t \ . \llabel{061116c} 
\ee

\subsubsection{Strong coupling}

We consider a strong coupling between the core and the mantle whenever $
\kappa^2 \gg \beta $, where $ \beta $ is given by expression (\ref{030123g}).
In this case, we can simplify expression (\ref{061116c}) by performing an
integration by parts:
\be
P (t) = \Frac{1}{\kappa} \Frac{P_\kk}{C} - \Frac{1}{\kappa^2} \Frac{C_m}{C^2}
\Frac{\partial P_\kk}{\partial t} \ , \llabel{061116d} 
\ee
where we neglected terms higher than $ \beta / \kappa^2 $.
Substituting the above equation (\ref{061116d}) into expression (\ref{050428b})
we find for the rotation rate variations:
\be
\Frac{d \omega}{d t} = \Frac{P_\kk}{C} + \Frac{T_\kk}{C} - \omega K_f \cos^2 \ve
\sin^2 \ve + \Frac{c_c}{\kappa} \Frac{\partial}{\partial t} \Frac{P_\kk}{C} \ .
\llabel{061116e}   
\ee

\subsubsection{Weak coupling}

For weak coupling we assume $ \kappa^2 \ll \beta $,  we will thus neglect
second order terms in $ \kappa^2 / \beta $.
Performing again an integration by parts in expression (\ref{061116c}), but
changing the roles of $ P_\kk / C_m $ and $ \ei^{\kappa_m t} $ we have:
\be
P (t) = \int \Frac{P_\kk}{C_m} \ d t \ , \llabel{061116f} 
\ee
which gives for the rotation rate variations when substituted into expression
(\ref{050428b}): 
\be
\Frac{d \omega}{d t} = \Frac{P_\kk}{C_m} + \Frac{T_\kk}{C} - \omega K_f \cos^2
\ve \sin^2 \ve - c_c \kappa_m \int \Frac{P_\kk}{C_m} \ d t \ .
\llabel{061116g}    
\ee

\section{Spin-orbit resonances}

\llabel{021023e}

The resonant equilibrium was first observed in the Moon, that is locked in a 1/1
spin-orbit resonance  \citep[e.g.][]{Goldreich_1966}.
That other spin-orbit resonances were possible was not realized before 
the discovery of the 3/2 spin-orbit resonance of Mercury
\citep{Pettengill_Dyce_1965}, giving rise to several detailed studies 
\citep{Colombo_1965,Colombo_Shapiro_1966,Goldreich_Peale_1966,Counselman_Shapiro_1970}.
Such non-synchronous spin-orbit resonances require a large orbital eccentricity,
but also, as we will see, low obliquity.

\subsection{Effect on the rotation rate}

Neglecting by now the dissipative effects resulting from tides and core-mantle
friction, when combining expressions (\ref{021014e}) and (\ref{060831a}), we can
write near a generic spin-orbit 
resonance $( \omega \simeq p n )$:  
\be
\frac{d \omega}{d t} = - \beta_m \sin 2 (\theta - p M - \phil)
\llabel{021016z} \ ,
\ee
with $ \beta_m = \betal / c_m $, where $ \betal $ and $
\phil $ are given in appendix~\ref{ApenB}, and $ c_m = C_m / C $.
Let us denote $ \gamma = \theta - p M - \phil $.
Since $ \omega = \dot \theta - \dot \psi \cos \ve $ we have
$ \dot \gamma = \omega - p n - \dot \phis $, with $ \dot \phis = \dot \phil +
\dot \psi \cos \ve $.
Because we assume $ \dot \psi \simeq \alpha \cos \ve $ (Eq.\ref{050214b}), 
for small variations of $ \omega $ and $ \ve $, we can consider $ \betal $, $
\dot \phil $ and $\dot \phis $ as constants.
Thus, we have $ \ddot \gamma = \dot \omega $ and expression (\ref{021016z}) can
then be rewritten as 
\be
\ddot \gamma + \beta_m \sin 2 \gamma = 0 \ , \llabel{061129g}
\ee 
which is the same as the equation of a free pendulum (Fig.\ref{Fig02}). 
The first integral of this equation is given by
\be
h = \dot \gamma^2 - \beta_m \cos 2 \gamma \ , \llabel{021014g}
\ee
where $ h $ is a constant of the motion related to the energy.
The separatrix equation is given by $ h = \beta_m $, where $ h > \beta_m $ gives
the trajectories in the circulation zone (outside the resonance) and 
$ h < \beta_m $ the trajectories in the libration zone (captured in resonance).
The maximal and minimal libration width, $ \omega_- \le \omega \le \omega_+ $, are
obtained from the separatrix equation ($ h = \beta_m $):
\be
\omega_\pm = p n \pm \Delta \omega \quad \mathrm{with} \quad
\Delta \omega = \sqrt{2 \beta_m} \llabel{021016x} \ ,
\ee
Since $ \beta_m \le \beta / c_m $ (Eq.\ref{030801z}), from expression
(\ref{030123g}) we have:
\be 
\frac{\Delta \omega}{n} \le \sqrt{3 \frac{B-A}{C_m}} \approx 0.02 \ ,
\llabel{021016y} 
\ee 
using $ (B-A)/C_m \simeq 1.2 \times 10^{-4} $ \citep{Anderson_etal_1987}.

\subsection{Dissipative torques}

\llabel{061122z}

Spin-orbit resonant configurations result from an evolutionary process. 
It is believed that the terrestrial planets' rotation was faster at the time
of their formation, but due to dissipative torques it has decreased
(section~\ref{021120w}) and may have been captured inside a resonance 
when crossing it. 

The secular variations of the rotation rate are easily computed from the
mantle's angular momentum variations (Eqs.\ref{050428b},\ref{061116b}) as:
\be
\Frac{d \omega}{d t} = \Frac{P_\kk}{C_m} + \overline D \ , \llabel{061122a}
\ee
where
\be
\overline D = \Frac{T_\kk}{C} - \omega K_f (1-x^2) x^2 - c_c \kappa_m P(t) \ 
\llabel{061122y}  
\ee
denotes the mean dissipative torque ($ x = \cos \ve $), which is 
composed of three terms: the first arising from tidal effects, the second
from non-resonant core-mantle friction and the last one is a dissipative term resulting from the
presence of spin-orbit resonances and core-mantle friction together.

For the tidal torque we can use the viscous model approximation
(section~\ref{021120a}), since significant spin-orbit resonances only occur in
the slow rotation regime ($ \omega \sim n $).
This torque is then given by expression (\ref{040811a}), which can be rewritten
using $ \omega = p n + \dot \gamma $ and $ E(e) = N(e)/\Omega(e) $ as
\be
\Frac{T_\kk}{C} = - K \Frac{1 + x^2}{2} \, \Omega(e) \left[ \left( p - \Frac{2
x E(e)}{1 + x^2} \right) + \Frac{\dot \gamma}{n} \right] \ . \llabel{061122x}
\ee

According to expression (\ref{021016y}) we have $ \dot \gamma / n \ll 1 $.
Thus, the non-resonant core-mantle friction torque can also be made linear:
\be
\omega K_f (\omega) 
\simeq p n  K_f (p n) \left( 1 + \Omega_f \Frac{\dot \gamma}{n} \right) 
\ , \llabel{061122w}
\ee
where
\be
\Omega_f = n \left[ \phantom{\frac{}{}} \! \! \ln (\omega K_f) \right]'_{\omega
= p n} = - \frac{q}{p} \ , \llabel{061122v}
\ee
and $ q $ is a semi-integer like $ p $. Indeed, since $ \alpha^2 \propto
\omega^{-2} $ (Eq.\ref{061120a}) and $ \kappa \propto \omega^{1/2} $
(Eq.\ref{V14a}), from expression (\ref{061122c}) we have for weak friction ($
\kappa \ll E_c \omega $) that 
\be
\omega K_f \propto \frac{1}{\omega^q} \ , \llabel{061129a}
\ee
with $ q = 5/2 $ and for strong friction ($ \kappa \gg E_c \omega $) the same
previous expression, but $ q = 3/2 $.
This result also works for turbulent friction, for which $ \kappa \propto 1 /
\omega $ and thus $ q = 4 $ \citep{Yoder_1995,Correia_etal_2003}.

Finally, for the resonant core-mantle friction contribution we will split our
analysis for strong and weak coupling as in section~\ref{061129c}.
In the first situation ($ \kappa^2 \gg \beta $) this calculus is trivial when
using expression (\ref{021016z}) for the precession torque.
Indeed, from expression (\ref{061116d}) we have
\begin{eqnarray}
- c_c \kappa_m P(t) & = & - c_c \Frac{P_\kk}{C_m} + \Frac{c_c}{\kappa}
\Frac{\partial}{\partial t} \left( \Frac{P_\kk}{C} \right) \crm
& = & c_c \beta_m \sin 2 \gamma - \frac{2 \betal}{\kappa} \dot \gamma \cos 2
\gamma \ . \llabel{061129d}
\end{eqnarray}
The term $ c_c \beta_m \sin 2 \gamma $ can be summed with the initial precession
torque (Eq.\ref{021016z}) providing a single term with amplitude $ \betal =
\beta_m - c_c \beta_m $. 
Thus, in the case of strong coupling the spin behaves as if there was
almost no differentiated internal structure, with only a small perturbation
resulting from the term in $ \cos 2 \gamma $.

In the case of weak coupling ($ \kappa^2 \ll \beta $), if we neglect the secular
effects (which is possible for short term variations), we can write using
equation (\ref{061129g}) 
\be
P(t) = \int \frac{P_\kk}{C_m} \, d t \simeq \int \ddot \gamma \, d t = \dot
\gamma - \dot \gamma_0 \ , \llabel{061129h} 
\ee
where $ \dot \gamma_0 $ is an integration constant.
This approximation is valid as long as the time interval $\Delta t$ for which we perform
the above integration verifies $ \Delta t \ll 1 / \kappa_m $
\citep{Correia_Laskar_2009}.

\subsection{Capture probabilities}

\llabel{041026z}

The total variation of the rotation rate when dissipative
torques are included is then
\be
\ddot \gamma + \beta_m \sin 2 \gamma = \overline D (\dot \gamma) \ .
\llabel{021014h} 
\ee
The spin-orbit term $ \beta_m \sin 2 \gamma $ is commonly known as the
restoration torque as it will counterbalance the dissipative torque $ \overline
D $ preventing the planet from escaping the resonant configuration.

\citet{Goldreich_Peale_1966} computed a first estimation of 
the capture probability $ P_\mathrm{cap} $,  
and subsequent more detailed studies proved their expression to be
essentially correct \citep[for a review, see][]{Henrard_1993}. 
We consider here the planet's orbit as a fixed ellipse at the moment it
approaches the resonance, since the perturbations of the orbital
parameters during this short period of time do not change the behavior of the
planet \citep{Goldreich_Peale_1966}.

Differentiating equation (\ref{021014g}) and replacing it in expression
(\ref{021014h}), we obtain
\be
 \frac{d h}{d t} = 2 \, \overline D (\dot \gamma) \, \frac{d \gamma}{d t} \ .
\ee 
The ``energy'' variation of the planet after a cycle around the
resonance is then given by the function: 
\begin{eqnarray}
\Delta h ( \gamma_1 ) & = & \int_{t_1}^{t_2} \frac{d h}{d t} \, d t + \int_{t_2}^{t_3}
\frac{d h}{d t} \, d t \\
& = & 2 \int_{\gamma_1}^{\gamma_2} \overline D (\dot \gamma) \, d \gamma + 2
\int_{\gamma_2}^{\gamma_3} \overline D ( \dot \gamma) \, d \gamma \ , \nonumber
\end{eqnarray}
where $ \gamma_1 $ is the $ \gamma $ value when the planet crosses the
separatrix between the circulation and the libration zones, i.e., the 
$ \gamma $ value for $ h = \beta_m $.
$ \gamma_2 $ and $ \gamma_3 $ are the first two following $ \gamma $ values
corresponding to $ \dot \gamma = 0 $.
$ t_1 $, $ t_2 $ and $ t_3 $ are the instants of time where the previous events
occurred, respectively.

In order to be captured, after a cycle inside the resonance, the
planet must remain within the libration zone.
When de-spinning from faster rotation rates, this means that the total ``energy''
of the planet after a cycle, $ h ( \gamma_3 ) $, must be smaller than the
separatrix ``energy'', $ h = \beta_m $ (Fig.\ref{Fig02}). 
Thus,
\be 
h ( \gamma_3 ) = h ( \gamma_1 ) + \Delta h ( \gamma_1 ) = \beta_m +
\Delta h ( \gamma_1 ) < \beta_m
\ee
and the capture condition becomes:
\be 
\Delta h ( \gamma_1 ) < 0 \ .
\ee

\prep{\FigB}

Assuming that the $ \gamma_1 $ values comprised between $ - \pi / 2 $ and $ \pi
/ 2 $ are distributed uniformly, capture inside the resonance will occur for
``energy'' variations within
\be
\left\{ {\cal E} : \Delta h ( - \pi / 2 ) < \Delta h ( \gamma_1 ) < 0 \right\}
\ee
from a total of possibilities:
\be
\left\{ {\cal E}_T : \Delta h ( - \pi / 2 ) < \Delta h ( \gamma_1 ) < \Delta h (
\pi / 2 ) \right\} \ ,
\ee
that is,
\be
P_\mathrm{cap}^+ = \Frac{\int_{\cal E} d \gamma_1}{\int_{{\cal E}_T} d \gamma_1} =
\frac{\Delta h ( - \frac{\pi}{2} )}{\Delta h ( - \frac{\pi}{2} ) - \Delta h (
\frac{\pi}{2} )} \llabel{021024k} \ .
\ee

Usually tidal torques are very weak, i.e., $ \md \overline D (\dot \gamma) \md
\ll \md \beta_m \md $, which gives $ \gamma_3 \simeq - \pi / 2 $,
$ \gamma_2 \simeq \pi / 2 $ and $ h \simeq \beta_m $. 
From expression (\ref{021014g}) we then write
\be 
\dot \gamma = \mathrm{sign} ( \dot \gamma ) \sqrt{2 \beta_m} \cos \gamma \ .
\llabel{090113a}
\ee
Replacing it in expression (\ref{021024k}) we compute for the capture
probability: 
\be
 P_\mathrm{cap}^+ = 1 + \Frac{\int_{\frac{\pi}{2}}^{-\frac{\pi}{2}} \overline D
 \left( \dot \gamma = - \sqrt{2 \beta_m} \cos \gamma \right) d
 \gamma}{\int_{-\frac{\pi}{2}}^{\frac{\pi}{2}} \overline D \left( \dot \gamma = 
+\sqrt{2 \beta_m} \cos \gamma \right)  d \gamma } \llabel{021024d} \ .
\ee

Following an identical reasoning, we can obtain the expression for the capture
probability when the spin is increasing from lower rotation rates:
\be
 P_\mathrm{cap}^- = 1 + \Frac{\int_{\frac{\pi}{2}}^{-\frac{\pi}{2}} \overline D
 \left( \dot \gamma = + \sqrt{2 \beta_m} \cos \gamma \right) d
 \gamma}{\int_{-\frac{\pi}{2}}^{\frac{\pi}{2}} \overline D \left( \dot \gamma = 
 - \sqrt{2 \beta_m} \cos \gamma \right)  d \gamma } \llabel{021030a} \ .
\ee

Let us notice that for even torques in $ \dot \gamma $, the capture never occurs, while for
odd torques it is unavoidable. 
The capture probability must lie between 0 and 1, but often the results given by
expressions (\ref{021024d}) and (\ref{021030a}) are outside this interval.
In those cases, if $ P_\mathrm{cap} < 0 $ then $ P_\mathrm{cap} = 0 $, and if $
P_\mathrm{cap} > 1 $ then $ P_\mathrm{cap} = 1 $.
For a general dissipation torque in the form 
\be
\overline D (\dot \gamma) = - K \left[ V + \left( \mu_1 + \mu_2 \cos 2 \gamma
\right) \frac{\dot \gamma}{n} \right] \ , \llabel{061204h}
\ee
where $ K $, $ V $, $ \mu_1 $ and $ \mu_2 $ are constants, we compute from
expressions (\ref{021024d}) and (\ref{021030a}):
\be
P_\mathrm{cap}^\pm = 2 \left[ 1 \pm \Frac{\pi}{2} \Frac{n}{\Delta \omega}
\Frac{V}{\mu} \right]^{-1} \quad \mathrm{with} \quad \mu = \mu_1 + 
 \frac{\mu_2}{3} \ . \llabel{061204i}
\ee

\section{Dynamical evolution}

\llabel{021030b}

In this section we analyze the dynamical equations obtained in previous sections.
The main goal is to describe both evolution and final stages for the spin
under the effect of dissipative effects.

\subsection{Rotation rate evolution}

\llabel{021120w}

The secular variations of the rotation rate are given by expression
(\ref{061122y}).
For a fast rotating planet, the spin is far from any spin-orbit resonance and we
can retain only the secular dissipative terms,
because $ \overline{P_\kk} = \overline{P(t)} = 0 $. 
In this regime we can use a constant-$Q$ model as good approximation
for tidal effects (Eq.\ref{040811r}).
Since all secular terms involved are negative (with $ \omega > 0 $), they can
only decrease the rotation rate for any value of the obliquity and eccentricity.
This is valid until the slow rotation regime is attained ($ \omega \sim n $),
where tidal effects can counterbalance the braking effect from core-mantle 
friction.
Once in this regime, two different behaviors are possible: the rotation rate
stabilizes around an equilibrium point given by the solution of $ T_\kk = 0 $
(section~\ref{021120c}) or the rotation rate is captured in a spin-orbit
resonance (section~\ref{021023e}).

\subsection{Obliquity evolution}

\llabel{040812a}

\subsubsection{Effect of core-mantle friction}

According to expression (\ref{061122b}), the secular effect of core-mantle
friction on the obliquity is given by:
\be
\Frac{d \ve}{d t} = - K_f \cos^3 \ve \sin \ve \ . \llabel{061122f}
\ee
Since $ K_f > 0 $ (Eq.\ref{061122c}), for any rotation rate the core-mantle friction always brings
the equatorial plane of the planet to the same plane as the orbit. 
We have that $ \dot \ve $ vanishes for $ \ve = 0^\circ $ and $ \ve =
180^\circ $ (which correspond to stable equilibrium positions) and for $ \ve =
90^\circ $ (unstable equilibrium).
Moreover, since $ K_f $ is proportional to $ \alpha^2 $ (Eq.\ref{061122c}) and $ \alpha \propto \omega^{-1} $
(Eq.\ref{061120a}), the magnitude of both $ \dot \omega $ and $ \dot \ve $ will 
grow as the planet slows down. 
Thus, for fast rotating planets the core-mantle friction effect can be
neglected, but as the planet arrives in the slow rotating regime ($ \omega \sim
n $), this effect grows so much that it may control the entire evolution of
the obliquity \citep{Correia_Laskar_2003I,Correia_etal_2003}. 
The decrease of the rotation rate and the obliquity variations are intimately
coupled. 
Indeed, combining the secular core-mantle friction contributions from
expressions (\ref{061122w}) and (\ref{061122f}), we have that $ \dot \omega \cos
\ve = \omega \dot \ve \sin \ve $ (the spin normal component is conserved). 
As a consequence, for an initial rotation rate $ \omega_{i} $ and obliquity $
\ve_i \ne 90^\circ $: 
\be 
\frac{\omega}{\omega_i} = \frac{\cos \ve_i}{\cos \ve}  \ . \llabel{eqANS} 
\ee
Since the obliquity evolves towards $ 0^\circ $ or $ 180^\circ$, i.e., $ \md
\cos \ve \md \rightarrow 1 $, the equilibrium rotation rate is
attained for $ \omega_e = \omega_i \, \md \cos \ve_i \md $.

\subsubsection{Effect of tides}

For a fast rotating planet, using again a constant-$Q$ model, the tidal effects
on the obliquity are given by the second equation in system (\ref{040811r}).
This equation has a single stable point for $ \ve = 67.11^\circ $ (or $ x =
0.388953 $).
Thus, since the core-mantle friction effect can be neglected in the fast
rotating regime (Eq.\ref{061122f}), whatever is the initial obliquity, it will 
evolve by tidal effect toward this balance point.

Once the planet arrives in the slow rotation regime ($ \omega \sim n $), the
constant-$Q$ model is no longer suitable and expression (\ref{040811r}) no
longer valid.  
Using the viscous model instead (Eq.\ref{040811a}), we find that the obliquity
still has only one stable point, obtained as the solution of $ d \ve / d t = 0 $:
\be 
\left\{ 
  \begin{array}{l l}
   \ve = \arccos \left( 2 n E(e) / \omega \right) 
   & \mathrm{if} \quad \omega > 2 n E(e) \ , \crm
   \ve = 0 & \mathrm{if} \quad 2 n E(e) \ge \omega > 0    \ ,
  \end{array} 
 \right.
 \llabel{V34}   
\ee
where $ E(e) = N(e) / \Omega(e) \ge 1 $ (Fig.\ref{Fig04}).
Contrary to the fast rotating regime, here the stable point
depends on the eccentricity and on the rotation rate of the planet.
The stable value for the obliquity decreases as the rotation slows down,
stabilizing for circular orbits at zero degrees for rotation rates of $ 2 n $ or
smaller, i.e., twice the orbital mean motion.
It results that tides, like core-mantle friction, always finish by bringing the
planet's equator to the orbital plane.
However, while core-mantle friction allows retrograde final rotations, tides
alone only admit direct rotations.

\subsection{Equilibrium positions}

\llabel{021120c}

True equilibrium positions for the spin result from simultaneous balance
points for the rotation rate and obliquity, that is
\be
\dot \omega = 0 \quad \mathrm{and} \quad \dot \ve = 0 \ .
\ee

In section~\ref{021120w} we saw that core-mantle friction and tidal effects both
decrease the rotation rate for a fast-rotating planet.
It is thus impossible to stabilize the spin before the planet arrives in the
slow rotation regime.
Once in this regime, core-mantle friction becomes dominant over tidal
effects \citep{Correia_etal_2003} and drives the obliquity into $ 0^\circ $ or $
180^\circ $ (Eq.\ref{061122f}). 
It follows then that we must look for stable values of the rotation rate
($ \dot \omega = 0 $) when $ x = \pm 1 $ in order to find the spin
equilibrium positions. 
For these obliquity values, in absence of the spin-orbit resonant term $ P_\kk
$, the contribution of the core-mantle friction to the rotation rate vanishes
(Eq.\ref{061122w}).
The equilibrium rotation rate is then determined solely by the tidal effects
(Eq.\ref{021010c1}), that is, when
\be
T_\kk = 0 \quad \Leftrightarrow \quad \sum_\sigma b (\sigma) \cQ^L_\sigma (\pm
1, e) = 0 \ , \llabel{040812b} 
\ee
or, using the viscous model (Eq.\ref{040811a}), simply
\be
\omega_e = E(e) \, n \ , \llabel{040812c}
\ee
which means that the equilibrium rotation rate increases
with the eccentricity of the planet \citep{Goldreich_Peale_1966,Hut_1981}.
This behavior is illustrated in Fig.\ref{Fig04}.

\prep{\FigC}

\subsection{Capture in resonance with $ \ve = 0^\circ $}

\llabel{theCase0}

In section~\ref{021120c} we saw that $ \ve = 0^\circ $ or $ \ve = 180^\circ $
correspond to the only two stable positions for the obliquity. 
For those obliquity values, the non-resonant core-mantle friction contribution
to the rotation rate vanishes (Eq.\ref{061122w}) and the rotation rate equation
(\ref{061122a}) can be greatly simplified.
Indeed, for $ \ve = 0^\circ $ ($x = 1$), we can write
\be
\ddot \gamma = - \Frac{\beta}{c_m} H (p, e) \sin 2 \gamma + \overline D_0 (\dot
\gamma) - c_c \kappa_m P(t) \ ,  \llabel{041026b}
\ee
where $ \gamma = \theta - p M - \phi $ and
\be
\overline D_0 (\dot \gamma) = - K \, \Omega(e) \left[  p - E(e) + \frac{\dot
\gamma}{n} \right] \ . \llabel{021024g}
\ee
A similar expression is obtained for the case $ \ve = 180^\circ $ ($ x = -1 $),
we just have to replace $ \omega $ by $ - \omega $ (see section~\ref{021205a}).
For a circular orbit $ (e = 0) $, we have $ H (p \neq 1, 0) = 0 $
(Tab.\ref{TAB1}) and
thus, the only spin-orbit resonance where capture can occur is the synchronous
resonance $ (p = 1) $. 
Capture in this resonance always occurs, because according to expression
(\ref{040812c}) we have $ \omega_e = n $, since $ E (0) = 1 $.

When the orbital eccentricity increases, according to expression
(\ref{040812c}) the equilibrium rotation rate $ \omega_e $ increases to
a larger value than $ n $ (Fig.\ref{Fig04}).
Since the synchronous resonance width (Eq.~\ref{021016y}) for $ \ve = 0^\circ $
is 
\be 
\Delta \omega \simeq n \sqrt{\frac{3 (B-A)}{C_m}} \ ,
\ee 
when $ \omega_e > n + \Delta \omega $, capture in this resonance becomes
impossible. 
The final rotation rate will then be given by $ \omega_e $ unless capture
in a resonance with $ p > 1 $ occurred.

\subsubsection{Absence of core-mantle friction}

In order to simplify, we will first compute the capture probabilities when there
is no contribution from the resonant core-mantle friction, i.e., we neglect the
term $ c_c \kappa_m P (t) $ in expression (\ref{041026b}).
Thus, replacing the expression of the tidal torque (Eq.\ref{021024g}) in
expression (\ref{061204i}), we get for the capture
probabilities in the $p$ resonance: 
\be
P_\mathrm{cap}^\pm \simeq 2 \left[ 1 - \frac{\varrho}{\mu} \pm \frac{1}{\mu}
\left( \frac{p - E(e)}{2 \Delta \omega / n \pi} \right) \right]^{-1}
\llabel{040820i} \ , 
\ee
with $ \varrho = 0 $ and $ \mu = 1 $.
This expression was first obtained by \citet{Goldreich_Peale_1966}.
It is straightforward that if $ (B-A)/C_m $ increases (or decreases), the capture
in resonance also increases (or decreases).
Using the present value for Mercury's inertia moment $ (B-A)/C_m \simeq 1.2
\times 10^{-4} $ \citep{Anderson_etal_1987} and $ e = 0.206 $, we compute for
the 3/2 spin-orbit resonance $ P_\mathrm{cap}^+ = 7.73 $\% and $ P_\mathrm{cap}^- = 0 $.
In figure~\ref{Fig05} we plotted several examples of capture in different
resonances as a function of the eccentricity when the planet is de-spinning from
faster rotation rates $( P_\mathrm{cap}^+ )$ and when its spin is increasing from
slower values $( P_\mathrm{cap}^- )$.
\prep{\FigD}
The most remarkable feature is that the probability grows very fast as the
eccentricity increases (or decreases for $ P_\mathrm{cap}^- $), but it suddenly
decays to zero. 
This important behavior was also described by \citet{Goldreich_Peale_1966} and
it corresponds to a zone where the equilibrium rotation rate does not reach
the resonance circulation zone, that is, when 
\be
\md \omega_e - p n \md > \Delta \omega \simeq n \sqrt{3 H (p, e)
\frac{B-A}{C_m}} \ . \llabel{061204c} 
\ee
Thus, the resonance width, which depends mainly on $ (B - A) / C_m $, not only
contributes for the probability of capture,
but also determines whenever this capture can occur or not.

\subsubsection{Weak coupling}

\llabel{061216d}

According to expression (\ref{061129h}), in the case of a weak coupling ($
\kappa^2 \ll \beta $) we can rewrite expression (\ref{041026b}) for $ \ve =
0^\circ $ as
\be
\ddot \gamma = - \Frac{\beta}{c_m} H (p, e) \sin 2 \gamma + \overline D (\dot
\gamma) \ ,  \llabel{061204e} 
\ee
where
\begin{eqnarray}
\overline D (\dot \gamma) & = & \overline D_0 (\dot \gamma) - c_c
\kappa_m (\dot \gamma - \dot \gamma_0) \crm & = & - K \, \Omega(e) \left[
 p - E(e) - \varrho \frac{\dot \gamma_0}{n} + \mu \frac{\dot \gamma}{n} \right] \ ,
 \llabel{061204f} 
\end{eqnarray}
with
\be
\varrho = \Frac{c_c \kappa_m \, n}{K \Omega(e)} \quad \mathrm{and} \quad \mu = 1
+ \varrho \ . \llabel{061204g} 
\ee
Just before the rotation rate enters the libration zone, its average value is
given by $ <\dot \gamma> = 2 \Delta \omega / \pi $ (Eq.\ref{090113a}).
In this regime, $ \kappa \ll \sqrt{\beta} \sim \Delta \omega $, and therefore
the core is unable to follow the periodic variations in the mantle's rotation
rate. 
Thus, if the planet is de-spinning from faster rotation rates, we can adopt $ \dot
\gamma_0 \simeq 2 \Delta \omega / \pi $ in expression 
(\ref{061204f}), since capture probabilities are computed when the rotation rate
crosses the separatrix between the libration and the circularization zones
\citep{Correia_Laskar_2009}. 
Likewise, if the planet is increasing its spin from slower rotation rates, we
use $ \dot \gamma_0 \simeq - 2 \Delta \omega / \pi $ in expression 
(\ref{061204f}).

In this case, the capture probabilities are also given by expression
(\ref{040820i}), with $ \varrho > 0 $. Since $ \mu = 1 + \varrho > 1 $, this
expression simplifies to $ P_\mathrm{cap}^\pm = 2 \mu / \left[ 1 \pm \left(
\frac{p - E(e)}{2 \Delta \omega / n \pi} \right) \right] $, that is, 
the capture probability is always higher than in absence of the resonant
contribution from core-mantle friction.
In figure~\ref{Fig05B}, we plotted several examples of capture in different
resonances as a function of the eccentricity when the planet is de-spinning from
faster rotation rates $( P_\mathrm{cap}^+ )$ and when its spin is increasing from
slower values $( P_\mathrm{cap}^- )$ using $ \nu = 10^{-6} \, \mathrm{m}^2
\mathrm{s}^{-1} $.
When comparing with figure~\ref{Fig05} (absence of core-mantle
friction), we observe that the capture probabilities largely increase, as
predicted by \citet{Peale_Boss_1977}.
For instance, with the present eccentricity of Mercury ($ e = 0.206 $) we get 
$ P_\mathrm{cap}^+ = 100 $\% of capture in the 3/2 resonance, which
contrasts with $ P_\mathrm{cap}^+ = 7.73 $\% in the absence of core-mantle
friction.
If we used a stronger value of the viscosity, $ \nu = 10^{-2} \, \mathrm{m}^2
\mathrm{s}^{-1} $, the capture probability in any of the four resonances plotted in
figures~\ref{Fig05} and~\ref{Fig05B} is always one if the eccentricity is
higher than $ 0.002 $ and smaller than $ 0.5 $, including for the 1/2 resonance.

\prep{\FigE}

\subsubsection{Strong coupling}

Following expression (\ref{061129d}), in the case of a strong coupling ($
\kappa^2 \gg \beta $) we can rewrite expression (\ref{041026b}) for $ \ve =
0^\circ $ as
\be
\ddot \gamma = - \beta H (p, e) \sin 2 \gamma + \overline D (\dot \gamma) \ , 
\llabel{061204b}  
\ee
where
\begin{eqnarray}
\overline D (\dot \gamma) & = & \overline D_0 (\dot \gamma) - \frac{2
\beta}{\kappa} H (p, e) \dot \gamma \cos 2 \gamma \crm & = & - K \, \Omega(e)
\left[ p - E(e) + \left( 1 + \mu_2 \cos 2 \gamma \right) \frac{\dot \gamma}{n} 
\right] \ , \llabel{061216a}
\end{eqnarray}
and
\be
\mu_2 = \frac{2 \beta}{\kappa} \frac{H (p, e) n}{K \Omega(e)} \ .
\llabel{061216b}
\ee
Thus, according to expression (\ref{061204h}), the capture probabilities are
still given by expression (\ref{040820i}) with $ \varrho = 0 $, $ C_m = C $, and
$ \mu = 1 + \mu_2 / 3 $.
Since $ \mu > 1 $, the consequences for the capture probability in resonance are
the same as in the weak coupling situation (section~\ref{061216d}).

\subsection{Capture in resonance with $ \ve \neq 0 $}

\llabel{theCaseNot0}

The final possible evolutions for the obliquity resulting from dissipative
effects are $ \ve = 0^\circ $ or $ \ve = 180^\circ $ (see
section~\ref{021120c}).
However, it may happen that when a resonance is crossed,
the obliquity is still evolving toward one of the final states.
It is then useful to analyze the consequences of a non-zero obliquity.
For simplicity, we will first look at the case of a planet without a core.
The complete effect with core-mantle friction is discussed in section~\ref{040819e}.

\subsubsection{Absence of core-mantle friction}

In absence of the core-mantle friction effect, for a non-zero obliquity value,
the equilibrium rotation rate $ \omega_e $ is obtained from expression
(\ref{040811a}) setting $ d L / d t = 0 $: 
\be
\omega_e = \Frac{2 x}{1+x^2} \, E(e) \, n \le E(e) \, n \ . \llabel{040819a} 
\ee
Thus, the main consequence of an increasing obliquity is to reduce the
equilibrium rotation rate (Fig.\ref{Fig06}).
For zero obliquity, the lowest equilibrium was the synchronous
motion ($ \omega_e = n $), so this configuration was the last possible
evolutionary stage for a planet de-spinning from faster spins. 
Now, if the spin of the planet is not captured in the 1/1 resonance, the
planet may evolve to slower final spin configurations, including the 1/2
spin-orbit resonance or below. 
Indeed, for $ \ve > 90^\circ $ the equilibrium is fixed at a negative rotation
rate, allowing the capture in negative resonances as well ($ p < 0 $).
Notice, however, that in this case the equilibrium spin always corresponds to
prograde rotation, since $ \ve > 90^\circ $.

\prep{\FigF}

The capture probabilities in all resonances will also vary for different
obliquities: 
not only the tidal torque is different (Eq.\ref{040811a}), but also the
resonance width will change (Eq.\ref{030731b}).
From equation (\ref{061204i}) we compute:
\be
P_\mathrm{cap}^\pm = 2 \left[ 1 - \frac{\varrho}{\mu} \pm \frac{1}{\mu} \left( p - \Frac{2 x
E(e)}{1+x^2} \right) \Frac{n \pi}{2 \Delta \omega} \right]^{-1} \ , \llabel{040819b}
\ee
where $ \varrho = 0 $, $ \mu = 1 $, and $ \Delta \omega = \sqrt{2 \beta_m} = \sqrt{2 \betal / c_m}
$, with $ \betal $ being given by expression (\ref{030731b}).
As the obliquity increases, the term in $ x $ will decrease and
the capture probability will then also decrease. 
In figure~\ref{Fig07} we plot the evolution of the capture probability with
the obliquity for the synchronous resonance and two different eccentricities
when the planet is de-spinning from faster rotation rates.
\prep{\FigH}
For $ e = 0 $ (circular orbit) and low obliquity the capture probability 
is 100\%. 
As the obliquity increases, this probability decreases fast, and it is close to
zero for obliquities higher than $ 90^\circ $. 
For eccentric orbits, we observe a sort of contrary effect: for some values of
the eccentricity, the obliquity can be responsible for an augmentation in the
capture probability.
Indeed, for high eccentricity and low obliquity the equilibrium
rotation rate is always faster than $ n $ (Fig.\ref{Fig04}) and
the capture in the synchronous resonance becomes impossible (Fig.\ref{Fig05}).
As the obliquity increases, the equilibrium rotation rate decreases
(Eq.\ref{040819a}) allowing the planet to approach again the synchronous
rotation rate. 
For $ e=0.2 $, this occurs when the obliquity is around $ 60^\circ $
(Fig.\ref{Fig07}).

\subsubsection{Capture in the 1/2 resonance}

\llabel{res12case}

An important consequence of a non-zero obliquity is that the spin can
reach previously impossible configurations (section~\ref{theCaseNot0}).
Because the capture in the 1/1 synchronous resonance becomes avoidable, the
equilibrium rotation rate can evolve to lower values than $ n $. 
In particular, capture in the 1/2 resonance can occur.
According to expression (\ref{040819a}), for a planet de-spinning from faster
rotation rates, unless $ \ve > 74.5^\circ $ the equilibrium rotation rate is
above the 1/2 resonance and the capture probability in this resonance is zero
(Eq.\ref{040819b}). 
Thus, there is only a chance of capturing the planet if the
resonance is crossed with high obliquity (Fig.\ref{Fig09}).

However, when the initial obliquity of the planet is high, following expression
(\ref{eqANS}) it can happen that the rotation rate becomes inferior
to $ n / 2 $ when the obliquity is brought to zero degrees by core-mantle
friction.
Then, the planet rotation rate will increase towards the equilibrium given by
equation (\ref{040812c}) and cross the 1/2 resonance from slower rotations.
The capture probability is given by $ P_\mathrm{cap}^- $ (Eq.\ref{040820i})
with $ p = 1/2 $.
Notice also that, as for any resonance different from the 1/1, capture in
the 1/2 resonance is only possible for a planet in an orbit with $ e > 0 $
(Tab.~\ref{TAB1}).
\prep{\FigI}

\subsubsection{Capture in ``negative'' resonances}

\llabel{negCase}

We now look at ``negative'' resonances, that is, resonances with $ p < 0 $
(Tab.\ref{TAB1}).
When $ \ve < 90^\circ $, we saw that the equilibrium rotation rate is always
positive (Eqs.\ref{eqANS} and~\ref{040819a}) and negative spin states
could not be attained.
However, for retrograde planets (whose obliquities are higher than $ 90^\circ
$), the equilibrium rotation rate is negative (Eq.\ref{040819a}).
Since capture in a ``positive'' resonance can hardly occur for these values of
the obliquity (Eq.\ref{040819b}), the rotation rate will decrease until the
planet starts to rotate in the opposite direction.
Once in this situation, the evolution for negative rotation rates can be
depicted from the positive rotations case. 
Due to the symmetry in the spin equations, the couple $ ( - \omega, \ve ) $
behaves identically to the couple $ ( \omega , \pi - \ve ) $. 
The capture probabilities can then be obtained directly from the positive
case: we just have to take into account that the situation corresponding to
capture from faster spin rates (slower in modulus) will now correspond to the
capture from slower spin rates and vice versa.
Contrary to the case with $ \ve = 0^\circ $, the $ 1/2 $
resonance will now be the first resonance to be encountered, followed by the
$ 1/1 $.

Capture in ``negative'' resonances is then a real possibility for planets
whose evolution leads the spin to the final obliquity $ \ve = 180^\circ $
(section~\ref{021120c}).
However, notice that, although $ p < 0 $, in this case the resonces always
correspond to prograde rotation, since $ \ve > 90^\circ $.
These spin-orbit resonances are thus not true negative resonances, and that is
why we wrote ``negative'' with quotes.
Indeed, an observer of the planet today is unable to determine if the 3/2
spin-orbit resonance corresponds to the spin state ($ \omega / n = 3 / 2 $, $ \ve
= 0^\circ $) or ($ \omega / n = - 3 / 2 $, $ \ve = 180^\circ $). 
A similar result had already been described for the spin of Venus 
\citep{Correia_Laskar_2001}, but for a different kind of equilibrium.

As a consequence, a true negative resonance will correspond to $p < 0$ and $\ve
< 90^\circ $ (or $p > 0$ and $\ve > 90^\circ $).
These resonant configurations are not impossible to achieve, but the capture
probability is very low (Figs.\ref{Fig07} and \ref{Fig08}) and we did not
register a single case during our numerical simulations.

\subsubsection{Core-mantle friction}

\llabel{040819e}


According to expression (\ref{061116g}) the core-mantle friction effect on the
spin has two different contributions, one permanent if the obliquity is
different from zero (Eq.\ref{061122w}) and another arising only near spin-orbit
resonances (Eq.\ref{061116c}).
This last contribution was already present for the case $ \ve = 0^\circ $ and
therefore examined in detail in section~\ref{061216d}.
As before, its contribution to the capture probabilities for a non-zero
obliquity is given by expression (\ref{040819b}) with, for weak coupling,
\be
\varrho = \frac{2 c_c \kappa_m n}{K (1+x^2) \Omega(e)} \quad \mathrm{and} \quad
\mu = 1 + \varrho \ , \llabel{061217z}
\ee
or, for strong coupling, $ \beta_m = \betal $,
\be
\varrho = 0 \quad \mathrm{and} \quad \mu = 1 + \frac{4 \betal n}{3 \kappa K
(1+x^2) \Omega(e)} \ . \llabel{061217x} 
\ee

The contribution of the non-resonant core-mantle friction term to the capture
probabilities is also easy to compute when we use the linear approximation given
by expression (\ref{061122w}).
In absence of any other dissipative effect, according to expression
(\ref{061204i}) it is given by
\be
P_\mathrm{cap}^+ \simeq - \, 4 \, \Frac{q}{p} \, \Frac{\sqrt{2 \beta_m}}{n
\pi} < 0 \ . \llabel{040820g}
\ee
This probability is always negative, which means that $ P_\mathrm{cap}^+ = 0 $
for any resonance. 
Because the spin can only decrease (Eq.\ref{eqANS}), we also have
$ P_\mathrm{cap}^- = 0 $.

In a more general case, tides are also present and must be taken into account.
The total dissipative torque can then be rewritten as
\be
\overline D (\dot \gamma) = - K \frac{1+x^2}{2} \Omega(e) \left[ \left( p -
\frac{2 x E(e)}{1+x^2} + \zeta \right) + \left( 1 - \zeta \frac{q}{p}  \right)
\frac{\dot \gamma}{n} \right] \ , \llabel{061218a} 
\ee
where
\be
\zeta = 2 p n \, \Frac{K_f(p n)}{K \Omega(e)} \, \Frac{\left( 1 -
x^2 \right)  x^2}{1+x^2} \ge 0 \llabel{090423a}
\ee
is the ratio between the magnitudes of the core-mantle friction and tidal
effects.
The total capture probability  is now obtained straightforwardly from expression
(\ref{061204i}) as
\be
P_\mathrm{cap}^\pm = 2 \left[ 1 - \frac{\varrho}{\mu_\zeta} \pm \frac{1}{\mu_\zeta}
\left( p - \Frac{2 x E(e)}{1+x^2} + \zeta \right)
\Frac{\pi n}{2 \Delta \omega} \right]^{-1} \ , \llabel{050516z}  
\ee
with 
\be 
\mu_\zeta = \mu - \zeta \Frac{q}{p} \llabel{081222a} \ , \ee
$ \varrho = 0 $ and $ \mu = 1 $.
If we want to take into account the resonant contribution from core-mantle
friction, we just need to modify $ \varrho $ and $ \mu $ according to
expressions (\ref{061217z}) and (\ref{061217x}).
For a dominating core-mantle friction ($ \zeta \gg 1 $) the previous
expression simplifies to expression (\ref{040820g}).
On the other hand, when $ \zeta \rightarrow 0 $ (weak core-mantle friction effect) we find
expression (\ref{040819b}).
This is always the case when the obliquity is close to $ 0^\circ $, $ 90^\circ $
or $ 180^\circ $, because $ x = 0 $ or $ x = \pm 1 $.

Since $ \zeta $ is always a positive quantity, expression (\ref{050516z}) shows
that the capture probability when de-spinning from faster spins 
is always smaller than it would be without the non-resonant core-mantle friction
($ \zeta = 0 $). 
In figure~\ref{Fig08} we plot the global effect for different spin-orbit
resonances and we observe an important reduction in the
capture probabilities as the obliquity increases.


\prep{\FigJ}

We then conclude that, contrary to the resonant contribution, the non-resonant
core-mantle friction effect is responsible for a reduction in the 
probability of capture for any resonance. 
Although the resonant core-mantle coupling greatly increases the chances of
capture in resonance at zero obliquity ($ P_\mathrm{cap}^+ = 100$\% for the 2/1
and 3/2 resonances), for high values of the obliquity this probability
considerably decreases, allowing the rotation of the planet to evolve into 
lower-order equilibrium configurations such as the 1/1 or the 1/2 resonances.

\section{Spin evolution}
\llabel{RTSM}

Using the dynamical equations derived in the former sections we will now
simulate the evolution of Mercury's spin.
We start our integrations shortly after the formation of the Solar System and
let the planet evolve until present days.
This strategy is needed because the present spin configuration of Mercury
corresponds to a stable equilibrium and it is impossible to remove it from this
state by simply reversing the time.

In order to proceed with our study we need to choose a set of plausible
coefficients for the dissipative models described in sections~\ref{030214b}
and~\ref{sectioCoreMantle}.
Although the contribution of these effects to the dynamical equations is well
understood, some of the geophysical parameters intervening are poorly known.
We will use a set of parameters called the standard model ($ k_2 = 0.4 $, $ Q =
50 $, $ \nu = 10^{-6} \, \mathrm{m}^2 \mathrm{s}^{-1} $), whose choice is
described in detail in \citet{Correia_Laskar_2009}. 

\subsection{Early evolution}

\llabel{OblEvol}

Before looking at the final stages of the spin evolution, where capture in
spin-orbit resonances may occur, we will study the behavior of the spin in the
early stages after planetary formation.
At this epoch Mercury is supposed to rotate much faster than today and any
orientation of its axis in space is allowed \citep{Dones_Tremaine_1993,Kokubo_Ida_2007}.
Indeed, the Caloris Basin, a large multi-ringed impact structure is estimated to
have been formed by the impact of a 150~km object about 3.85~Gyr ago,
at the end of the period known as late heavy bombardment
\citep{Murdin_2000,Strom_etal_2008}. 

Because Mercury is believed to spin rapidly at the beginning of its evolution ($
\omega \gg n $), a constant $ Q $ model (section~\ref{021024j}) seems to be the
best choice for tidal evolution.
However, for the present slow rotation ($ \omega \sim n $), a viscous tidal
model is the most appropriate.
In our simulations we then interpolate between the two models,
adopting $ Q = 20 $ for a fast rotating planet with a constant
dissipation model, and $ Q = 50 $ in the limit of slow rotations with a viscous
dissipation model. The transition is done at $ \omega \sim 10 \, n $. 

\prep{\FigK}

We follow the spin evolution for an initial rotation
period of about 32~h ($ \omega = 65 \, n $) and different initial obliquities
spanning from $ 0^\circ $ to $ 180^\circ $ (Fig.\ref{FM01}).
Two distinct behaviors are observed: for initial obliquities smaller than about 
$ 175^\circ $ the obliquity is brought to zero. For higher initial obliquities
the final obliquity is $ 180^\circ $ and the final rotation rate is negative.
As discussed in section~\ref{040812a}, there are only two final
possibilities for the obliquity.
The bifurcation in these two distinct evolutions is provoked by core-mantle
friction. 
In the fast rotating regime ($ \omega \gg n $), this dissipative effect can be
neglected (Eq.\ref{061122c}) and tidal effects drive the obliquity to the
equilibrium value $ \ve \simeq 67^\circ $ (Eq.\ref{040811r}).
This is why initial obliquities lower than $ 67^\circ $ increase while higher
decrease. 
Once in the slow rotation regime ($ \omega \sim n $), the tidal equilibrium
obliquity will move toward zero (Eq.\ref{V34}).
However, in this regime, core-mantle friction becomes stronger than tidal effects and
drives the final obliquity evolution. 
When core-mantle friction becomes dominant, if the obliquity is still higher
than $ 90^\circ $ it will then end at $ 180^\circ $ (Eq.\ref{061122f}).
In absence of core-mantle friction the obliquity would always end at zero degrees.

For a stronger core-mantle friction effect, the global picture in figure~\ref{FM01}  would not
change much, only the critical initial obliquity that triggers the two distinct
final evolutions would be lower than $ 175^\circ $.
For instance, using $ \nu = 10^{-4} \, \mathrm{m}^2 \mathrm{s}^{-1} $ this
threshold drops to $ 170^\circ $. 
It is also important to note that the critical obliquities increase for faster
initial rotation rates and decrease for slower ones.


\subsection{Final evolution}

We now look in detail at the behavior of the spin at the final stages of its
evolution. 
Here the planet will encounter several spin-orbit resonances where the rotation
rate may be trapped.
Since we are considering the effect of core-mantle friction, we need to take
into account its resonant contribution to the rotation rate (Eq.\ref{061116c}),
as discussed in section~\ref{061122z}. 
With the presently known geophysical parameters of Mercury, using the core
viscosity $ \nu = 10^{-2} \, \mathrm{m}^2 \mathrm{s}^{-1} $, 
we compute $ \kappa^2 / \beta \sim 10^{-2} \ll 1 $,
the $ \kappa^2 / \beta $ ratio being smaller for lower viscosity values.
Thus, we conclude that we can use the weak coupling approximation
(Eq.\ref{061129h}) for Mercury, and the evolution of its rotation rate can be
described by expression (\ref{061204e}).
Nevertheless, in our simulations we will integrate simultaneously the mantle
(Eq.\ref{050428b}) and the core (Eq.\ref{061116z}) rotation rates.

As we just saw in previous section, the final evolution of the obliquity is
either along $ 0^\circ $ or $ 180^\circ $. 
We will then first consider the simplified situation where the obliquity already
achieved one of the final states ($ \ve = 0^\circ $) when the rotation rate
enters the zone of spin-orbit resonances.
This will allow us better to compare our results with those from previous
studies, for which the obliquity was always held fixed at zero degrees.
Because there is no guarantee that at the time of the first resonance crossing
the obliquity is already close to zero (Fig.\ref{FM01}) and
since the capture probabilities also change with the obliquity
(section~\ref{theCaseNot0}), we will then look at the evolution of the rotation
rate when the obliquity is still varying.

\subsubsection{Case $ \ve = 0^\circ $}

Once the obliquity reaches $ 0^\circ $ (or $ 180^\circ $), the non-resonant
effect of core-mantle friction vanishes (Eqs.\ref{061122w},\ref{061122f}).
Tidal dissipation will then drive the rotation rate of the planet 
towards a limit equilibrium value $ \omega_e $ depending on the eccentricity
$e$ and on the mean motion $n$ (Eq.\ref{040812c}).
In a circular orbit ($ e = 0 $) this equilibrium coincides with synchronization
($ \omega_ e / n = 1 $), while the equilibrium rotation rate 
$ \omega_e / n = 1.5 $ is achieved for $ e_{3/2} = 0.284927 $ (Fig.\ref{Fig04}). 
For the present value of  Mercury's eccentricity ($ e = 0.206 $) we have $
\omega_e / n = 1.25685$.
Thus, when Mercury is de-spinning from faster rotation rates it will encounter
all spin-orbit resonances with $ p \ge 3/2 $, and when the spin is increasing
from lower values it can be captured in resonances with $ p \le 1/1 $.
In absence of planetary perturbations, the eccentricity remains constant and each
resonance is crossed only once.
In order to estimate numerically the probability of capture $ P^+_\mathrm{cap} $,
we kept the initial rotation period of 32~h, and ran 2000 simulations for an
initial obliquity of $ 0^\circ $ with only slightly different initial libration
phase angles. 
Since the resonant part of the core-mantle friction contribution
(Eq.\ref{061116c}) modifies the probability of capture (section~\ref{theCase0}),
we performed our experiments first in absence of core-mantle friction, and then
including this effect with $ \nu = 10^{-6} \, \mathrm{m}^2 
\mathrm{s}^{-1} $. 
Results are shown in Table~\ref{TBNat3}.

\prep{\TabB}

In absence of core-mantle friction and using the present eccentricity of Mercury
($ e = 0.206 $), the probability of capture in the 3/2 
spin-orbit resonance was numerically estimated to be 7.2\%. 
In their work, \citet{Goldreich_Peale_1966} estimated analytically the
same probability to be $ P_{3/2} =$ 6.7\% (Eq.\ref{040820i}). 
With the updated value of the moment of inertia $ (B-A)/C_m \simeq 1.2 \times
10^{-4} $ \citep{Anderson_etal_1987}, this probability is increased to
7.7\%, which is in a satisfactory agreement with our numerical simulations.
The same is also true for all the other resonances (Tab.\ref{TBNat3}). 

When we add the effect from core-mantle friction, the numerical simulations are
still in a good agreement with the theoretical estimations given by expression
(\ref{040820i}), showing that they can be used to forecast the behavior of
numerical simulations.
In this situation the capture probabilities considerably increase for
all spin-orbit resonances. In particular, capture probability in the 2/1 resonance
also becomes 100\%, preventing a subsequent evolution to the 3/2 resonance.
This behavior was already expected from our analysis in section~\ref{061216d}, 
and it is in conformity with the results from \citet{Goldreich_Peale_1967}
and \citet{Peale_Boss_1977}, that is, when the effect from core-mantle
friction is considered, the probabilities of capture are greatly enhanced.

\subsubsection{Case $ \ve \ne 0^\circ $}

The initial obliquity of Mercury is unknown, since a small number of large
impacts at the end of the formation process will not average away and may
change the obliquity of the planet \citep{Dones_Tremaine_1993,Kokubo_Ida_2007}.
In addition, even for initial low obliquities, during the first stages of the
evolution, the strong tidal effects acting on Mercury tend to increase the
obliquity (Fig.\ref{FM01}).
Thus, when the planet arrives in the slow rotation regime ($ \omega \sim n $),
for which resonance crossing occurs, it is almost certain that the obliquity is
higher than zero.

According to expression (\ref{040819a}) as the obliquity increases, the
equilibrium rotation rate decreases (Fig.\ref{Fig06}) allowing the spin to
evolve into spin-orbit resonances lower than the synchronous rotation ($ p = 1
$).
This possibility is also enhanced by the fact that the capture probability in
positive resonances ($ p > 0 $) is also reduced for high obliquities
(Figs.\ref{Fig07} and~\ref{Fig08}).
Thus, when considering a non-zero obliquity we may expect the planet to evolve
into resonant configurations with $ p < 3/2 $.

To test these scenarios, we repeated the previous 2000 numerical experiences
using several different initial obliquities with $ \nu = 10^{-6} \, \mathrm{m}^2
\mathrm{s}^{-1} $. 
In Table~\ref{TM01} we report the distribution of the different final spins
obtained.
As expected, we observe a significant
modification in the number of captures for all resonances, and for retrograde
planets these captures can occur in lower-order resonances than the 3/2.

\prep{\TabC}

For initial obliquities lower than the critical obliquity
($ \ve_0 \approx 175^\circ $ for $ \nu = 10^{-6} \, \mathrm{m}^2 \mathrm{s}^{-1}
$) 
the capture probability in all resonances is reduced
according to expression (\ref{050516z}). 
Indeed, when the resonances are crossed with non-zero obliquity, not only the
capture probability resulting from tides is smaller, but there is also a
reducing contribution from the core-mantle friction effect (Fig.\ref{Fig08}). 
For instance, 
when the initial obliquity is $ \ve_0 = 90^\circ $, the 2/1 resonance is
crossed with an obliquity around $ \ve \simeq 50^\circ $, and
capture in the 3/2 resonance becomes possible, while it was not for $ \ve
= 0^\circ $ (Fig.\ref{Fig08}).
The fraction of captures in this resonance presents some variations between 1\% and
7\%, because it depends on the probability of not being captured in higher-order 
resonances (Tab.\ref{TM01}).


For initial obliquities higher than $170^\circ$ (but still lower than the
critical obliquity $ 175^\circ $), we observe that
capture in resonances lower than the 3/2 also occurred. 
The reason is that, as core-mantle friction decreases the obliquity, it
also decreases the rotation rate following expression (\ref{eqANS}).
When the obliquity reaches zero, the rotation rate may be lower than
the equilibrium rotation rate determined by tides (Eq.\ref{040812c}), i.e., $ 0
< \omega < E(e) n $, with $ E(0.206) = 1.25685 $.
Then, tides alone will increase the spin toward the equilibrium position, and if $
\omega $ is below the spin-orbit resonances 1/2 or 1/1, capture in these
resonances can occur. 
Notice however, that in this case the capture probabilities are given by the
expression of $ P^-_\mathrm{cap} $, since the planet crosses the resonance when
the spin is increasing from lower rotation rates.
This is why capture in the 1/2 resonance is possible (Fig.\ref{Fig05B}).

For initial obliquities higher than the critical value, the obliquity of the
planet is always higher than $ 90^\circ $ when the resonances are crossed
(Fig.\ref{FM01}).
Thus, the probability of capture in all ``positive'' resonances ($ p > 0 $) 
is very small (Fig.\ref{Fig08}).
In this situation the equilibrium rotation rate given by tides
will be set at $ \omega = - 1.25685 \, n $ (Eq.\ref{040819a}).
A planet with $ \omega > 0$ will continue to decelerate, skip all the
``positive'' resonances, reverse its rotation direction, and
accelerate its spin rate again.
It will then sequentially cross resonances $ 1/2 $ and $ 1/1 $ until it
reaches $ \omega = - 1.25685 \, n $, if not captured before in one of those two
resonances.
The capture probabilities in these two ``negative'' resonances are the same as for
the positive resonances $ 1/2 $ and $ 1/1 $ for a planet with $ \ve = 0^\circ $
when increasing its spin from lower values (section~\ref{negCase}).
In the case with $ \nu = 10^{-6} \, \mathrm{m}^2 \mathrm{s}^{-1} $, we have a
31.9\% chance of capture in the 1/2 resonance and 100\% in the 1/1, that is,
all the simulations that avoided the 1/2 resonance were then trapped in the 1/1
(Tab.\ref{TBNat3}).




\section{Conclusions}

In the present work we derived a formalism to describe the complete evolution of
the spin of a terrestrial planet like Mercury under the effect of strong solar tides and core-mantle friction.
The inclusion of the obliquity in the equations of motion allowed us to compute
the capture probabilities in resonance for any initial spin value.
Even though zero obliquity is the final equilibrium resulting from dissipative
effects, for many sets of initial conditions it is possible that Mercury
encounters a resonance with a large obliquity.
As we increase the obliquity of the planet, the probability of capture in
resonance always decreases, allowing the spin to evolve to unexpected
configurations.

The presence of a liquid core and the associated core-mantle friction effect
also may lead to a peculiar evolution of the spin.
Indeed, if the obliquity is higher than $ 90^\circ $ at the moment this effect becomes
dominant over tides, the final obliquity will be set at $ 180^\circ $.
In this case the rotation rate evolves into negative values, and
the final spin is always prograde. 

Another important consequence of core-mantle friction is to produce significant
modifications of the capture probabilities in resonance. 
This effect can be decomposed in two, one resulting from the libration of the
mantle near spin-orbit resonances (resonant contribution) and another resulting
from different orientations of the core and the mantle spin vectors when the
obliquity is not zero (non-resonant contribution).
This last effect leads to a decrease in the capture probabilities in resonance,
while the first one increases those chances.

Finally, we performed some numerical simulations for Mercury, starting with a
fast rotating planet and different obliquity values.
This allowed us to study the final distribution possibilities for the rotation
rate.
Higher-order resonances than the 3/2 were already expected, since Mercury
had to cross them when de-spinning from faster rotation rates. 
However, we could also observe lower-order resonant configurations such as the
synchronous or the 1/2 spin-orbit resonance. 
For retrograde planets, ``negative'' resonances ($ p < 0 $) were also observed, but
also corresponding to prograde final states, since $ \ve = 180^\circ $.
The present formalism and results should apply more generally to any extrasolar planet or satellite
with a core and whose evolution led to cross spin-orbit resonances.

The synchronous resonance can also be achieved for zero obliquity if the chaotic
evolution of the eccentricity is taken into account,
when very low values of the eccentricity destabilize higher-order resonances 
and drive the planet's rotation towards the 1/1 resonance
\citep{Correia_Laskar_2004,Correia_Laskar_2009}. This 
cannot occur  for the present value of the eccentricity ($ e = 0.206 $).
In a forthcoming study we will include the contribution 
of planetary perturbations, which will require massive numerical simulations. 
This will add the contribution of the variation of eccentricity,
already taken into account by \citet{Correia_Laskar_2009}, 
but in addition, using the formalism that has been developed 
in the present work, we will be able also to take into account 
the large chaotic variations of the planet's obliquity 
resulting from planetary perturbations \citep{Laskar_Robutel_1993}.



\section*{Acknowledgments}

The authors thank S.J. Peale for discussions.
This work was supported by the Funda\c{c}\~{a}o para a Ci\^{e}ncia e a
Tecnologia (Portugal) and by PNP-CNRS (France).

\appendix

\section{The mean potential energy}

\llabel{ApenB}

In section~\ref{030123a} we eliminate the fast angles $ \theta $ and $ v $ from
the potential energy $ \cU $ by expanding
the true anomaly $ v $ in series of the mean anomaly $ M $ and then by
taking the average of $ \cU $ over $ \theta $ and $ M $.
However, resonant terms with argument $ (\theta - p M) $ appear in the
expression of the potential energy (Eqs.~\ref{030123b}, \ref{030123c}) that must
be taken into account to   
the mean   potential energy $ \overline \cU $ (Eq.~\ref{030123e}).
Here we give the derivation of the amplitudes $ \betal $ and $ \beta_r $ as well
as the respective phase angles $ \phil $ and $ \phi_r $.

Let $ \cU_r $ be the resonant part of the potential energy (Eq.~\ref{030123b})
\be 
\frac{8 \, \cU_r}{C} = - 2 \, \beta \left( \frac{a}{r_{}} \right)^3 F (\theta, w,
\ve) \; , \llabel{050602z}
\ee
where $ \beta $ is given by expression (\ref{030123g}).
We can rewrite $ \cU_r $ as the real part of
\begin{eqnarray}
\frac{8 \, \hat \cU_r}{C} & = & - \beta \left( \frac{a}{r_{}} \right)^3 \left[
2 (1-x^2) \ei^{\ii 2 \theta} \phantom{\frac{1}{1}} \right. \crm & & \left.
\phantom{\frac{1}{1}} + (x+1)^2 \ei^{\ii (2 \theta - 2 w)} + (x-1)^2 \ei^{\ii (2
\theta + 2 w)} \right] \ , \llabel{061120f}
\end{eqnarray}
and, averaging over $ \theta $ and $ w $ using expressions (\ref{061120ga}) and
(\ref{061120gb}),
\begin{eqnarray}
\frac{8 \, \hat \cU_r}{C} & = & - \beta \sum_{p = - \infty}^{\infty} \left\{ 2 (1-x^2) G (p, e) \ei^{\ii 2
(\theta - p M)} \right. \crm
& &  \quad + \quad H ( p, e) \left[
(x+1)^2 \ei^{- \ii 2 \phi} \ei^{\ii 2 (\theta - p M)} \right. \crm
& & \hspace{2.2cm} \left. \left. + (x-1)^2 \ei^{\ii 2 \phi}
\ei^{\ii 2 (\theta + p M)} \right] \right\} \ . \llabel{030731a}
\end{eqnarray}
where $ \phi = \varpi + \psi $. 
Let
\be
g  = \frac{1}{4} (1-x^2) G (p, e) \quad \mathrm{and} \quad h^\pm  = \frac{1}{8} 
(x \pm 1)^2 H
( \pm p, e) \ . \llabel{030801y}
\ee
Retaining only the the resonant terms with argument $ (\theta - p M) $ in
expression (\ref{030731a}), we rewrite expression (\ref{050602z}) simply
as: 
\be
\frac{\hat \cU_r}{C} = - \underbrace{\beta \left[ g  +
h^+ \ei^{- \ii 2 \phi}  + h^-  \ei^{\ii 2 \phi}
\right]}_{= \ \betal \, \ei^{- \ii 2 \phil}} \ei^{\ii 2 (\theta - p M)} \ ,
\llabel{030731d}
\ee
where
\begin{eqnarray}
\left( \Frac{\betal}{\beta} \right)^2 & = & \left( g + h^+ + h^- \right)^2 - 4 g
\left( h^+ + h^- \right) \sin^2 \phi \crm
& & - 4 h^+ h^- \sin^2 2 \phi 
\llabel{030731b}
\end{eqnarray}
and
\be
\tan 2 \phil = \Frac{(h^+ - h^-) \sin 2 \phi}{g + (h^+ + h^-) \cos 2
\phi} \ . \llabel{030731c}
\ee
If $ x = 1 $, we have $ \betal = \beta H(p,e) $ and $ \phil = \phi $.
Notice that if $ g, h^\pm \ge 0 $ (which is often the case), we have
\be 
\betal \le \beta \, ( g + h^+ + h^- ) \ , \llabel{030801z}
\ee
and also that the average value over $ \phi $ is:
\be
\left< \left( \frac{\betal}{\beta} \right)^2 \right>_\phi = g^2 + (h^+)^2 +
(h^-)^2 \ .
\ee

The amplitude $ \alpha_r $ and the phase angle $\phi_r $ can be similarly
obtained from expression (\ref{030731d}).
Indeed, from expression (\ref{050602a}) we have:
\begin{eqnarray}
\frac{d \ve}{d t} & = & - \Frac{1}{8 \omega \sin \ve}\left[ x \frac{\partial}{
\partial \theta} + \frac{\partial}{\partial \psi} \right] \left( \frac{8 \, \hat
\cU_r}{C} \right) \crm & = & - \underbrace{\Frac{\beta}{\omega} \left[ g_r +
h^+_r \ei^{- \ii 2 \phi}  + h^-_r  \ei^{\ii 2 \phi}
\right]}_{= \ \alpha_r \, \ei^{-\ii 2 \phi_r}} \ii \sin
\ve \, \ei^{\ii 2 (\theta - p M)} \ , \llabel{090419a} 
\end{eqnarray}
where
\be
g_r = - \frac{x}{2} G (p, e) \quad \mathrm{and} \quad
 h^\pm_r = \frac{1}{4} (x \pm 1) H (\pm p, e) \ . \llabel{090423b}
\ee
The expressions for $ \omega \, \alpha_r $ and $ \phi_r $ are respectively
given by the expressions for $ \betal$ and $ \phil $ (Eqs.~\ref{030731b},
\ref{030731c}), where the quantities $ g $ and $ h^\pm $ are respectively
replaced by $ g_r $ and $ h^\pm_r $.

\section{Determination of the expressions for the functions $ \Omega
(e) $ and $ N (e) $}

\llabel{ApenC}

In section~\ref{021120a} we wrote the spin equations of motion
(Eqs.~\ref{040811a}) under the effect of tides for a viscous
dissipation model. 
For $ \ve = 0 $, these equations were already obtained by
\citet{Goldreich_Peale_1966} and \citet{Hut_1981}.
Here we will derive those expressions for any value of the eccentricity
and obliquity. 
We already used them in \citet{Levrard_etal_2007}, without demonstration.
Using the same notation as in sections~\ref{030214c} and~\ref{021205a}, we
rewrite the tidal potential (Eq.~\ref{021010a}) as:
\be
\cV^g = -k_2 \Frac{G m_\odot R^5}{r_{}^3 r'^3} P_2 ( \cos S ) \ ,
\ee
with $ \cos S = \vv{\hat r} \cdot \vv{\hat r}' $, the prime $ ' $ referring
to the interacting body. 
Assuming that both interacting and perturbing body are in the same orbital
plane, we can write 
\begin{eqnarray}
\cos S & = & \Frac{(1+x)^2}{4} \cos ( w - w' - \theta + \theta' ) \crm & + &
\Frac{(1-x)^2}{4} \cos ( w - w' + \theta - \theta' ) \crm  
& + & \Frac{1-x^2}{2} \left[ \cos ( w - w' ) \phantom{\Frac{}{}} \right. \crm
&  & \hspace{1cm} + \left. \cos ( w + w' ) \left( \cos (
\theta - \theta' ) - 1 \phantom{\Frac{}{}} \right) \right] \ , \llabel{050712a}
\end{eqnarray}
where $ w' = \varpi' + \psi + v' $ is the true longitude of date.
It is now easy to evaluate the contributions to the spin variations using
equations (\ref{021010b}).
For the variation of the rotation rate we obtain:
\be
\frac{d L}{d t} = k_2 \Frac{3 G m' m_\odot R^5}{r_{}^3 r'^3} \, \cos S
\, \Frac{\partial \cos S}{\partial \theta} \ . \llabel{030217c}
\ee
Let $ \Delta t $ be the time delay between the perturbation and the planet's
response.
Then, assuming $ \Delta t $ small and the interacting body the
same as the perturbing one ($ m' = m_\odot $), we write
\citep{Mignard_1979,Mignard_1980}:
\be
\theta' = \theta (t - \Delta t) \simeq \theta (t) - \frac{d \theta}{d t} \Delta t
\simeq \theta - \omega \Delta t \llabel{030217a}
\ee 
and
\be
v' = v (t - \Delta t) \simeq v (t) - \frac{d v}{d t} \Delta t 
= v - n \frac{a^2}{r_{}^2} \sqrt{1 - e^2} \Delta t \ . \llabel{030217b}
\ee 
Substituting the above expressions (\ref{030217a}) and (\ref{030217b}) into
expression (\ref{030217c}) for the rotation rate, we have to first order in $
\Delta t $:
\begin{eqnarray}
\frac{d L}{d t} & \simeq & - k_2 \Frac{3 G m_\odot^2 R^5}{r_{}^6} \, \left[
\left( \Frac{1+x^2}{2} + \Frac{1-x^2}{2} \cos (2 w) \right) \omega \right. \crm 
& & \hspace{2.5cm}  - \left. \left( x \,
\frac{a^2}{r_{}^2} \sqrt{1 - e^2} \right) n \right] \Delta t \ . 
\end{eqnarray}
Using $ a / r = (1 + e \cos v) / (1 - e^2) $ in the previous expression and
averaging it over the mean anomaly $ M $ and the longitude of the perihelion $
\varpi $, we finally get:
\be
\frac{d L}{d t} = - \Frac{G m_\odot^2 R^5}{a^6} \, \Frac{3 k_2}{Q} \, 
\left[ \left( \Frac{1+x^2}{2} \right) \Omega (e) \,\frac{\omega}{n} - x \, N (e)
\right] \ ,  
\ee
where $ Q^{-1} = n \Delta t $,
\be
\Omega (e) = \Frac{1 + 3 e^2 + 3 e^4 / 8}{(1 - e^2)^{9/2}} 
\ee
and
\be 
N (e) = \Frac{1 + 15 e^2 / 2 + 45 e^4 / 8+ 5 e^6 / 16}{(1 - e^2)^{6}}  \ .
\ee

\section{Nomenclature}

\llabel{ApenD}

\begin{table}
\begin{tabular}{|c|l|c|} \hline
 {\it Symbol}  & $ 
\quad \quad \quad \quad \quad $ Designation $ 
\quad \quad \quad \quad \quad \quad $  & Eq. \\  \hline
$ a $ & Mercury's semi-major axis &  \ref{061120ga} \\
$ A $ & minimal moment of inertia &  \ref{090418a} \\
$ A_c $ & core's minimal moment of inertia &  \ref{050514a} \\
$ b (\sigma) $ &  tidal dissipation factor & \ref{021010e} \\
$ B $ & intermediate moment of inertia &  \ref{090418a} \\
$ c_c $ & core's  moment of inertia ($c_c = C_c/C$) &  \ref{061122c} \\
$ c_m $ & mantle's  moment of inertia ($c_m = C_m/C$) &  \ref{021016z} \\
$ C $ & maximal moment of inertia &  \ref{090418a} \\
$ C_c $ & core's maximal moment of inertia &  \ref{050514a} \\
$ C_m $ & mantle's maximal moment of inertia &  \ref{060831a} \\
$ \overline D $ & general mean dissipative torque &  \ref{061122y} \\
$ e $ & eccentricity of Mercury's orbit &  \ref{061120ga} \\
$ E $ & total tidal energy &  \ref{090419e} \\
$ E (e) $ & tidal eccentricity function & \ref{061122x} \\
$ E_c $ & core dynamical ellipticity &  \ref{050514a} \\
$ E_d $ & dynamical ellipticity &  \ref{050109a} \\
$ g $ & function depending on $x$ and $G(e,p)$ &  \ref{030801y} \\
$ g_r $ & function depending on $x$ and $G(e,p)$ &  \ref{090423b} \\
$ G $ & gravitational constant &  \ref{090418a}  \\
$ G (p,e) $ & power series in $e$ &  \ref{061120ga} \\
$ h $ & constant of motion related to the energy &  \ref{021014g} \\
$ h^\pm $ & function depending on $x$ and $H(e,p)$ &  \ref{030801y} \\
$ h_r^\pm $ & function depending on $x$ and $H(e,p)$ &  \ref{090423b} \\
$ H (p,e) $ & power series in $e$ &  \ref{061120gb} \\
$ \vv{\ii} $ & minimal axis of inertia &  \ref{090418a} \\
$ \vv{\mathrm{I}} $ & reference axis of inertial frame &  \ref{090418b} \\
$ \tilde {\cal I}_c $ & core tensor of inertia &  \ref{050514a} \\
$ \vv{\jj} $ & intermediate axis of inertia &  \ref{090418a} \\
$ \vv{\mathrm{J}} $ & reference axis of inertial frame &  \ref{090418b} \\
$ \vv{\kk} $ & maximal axis of inertia &  \ref{090418a} \\
$ \vv{\mathrm{K}} $ & normal axis to the ecliptic plane &  \ref{090418b} \\
$ k_2 $ & second Love number &  \ref{021010a} \\
$ k_f $ & fluid Love number &  \ref{050109a} \\
$ K $ & tidal dissipation amplitude &  \ref{eq3} \\
$ K_f $ & core-mantle friction amplitude  & \ref{061122c} \\
$ \vv{L} $ & total angular momentum &  \ref{060113a} \\
$ \vv{L}_c $ & core angular momentum &  \ref{050514a} \\
$ \vv{L}_m $ & mantle angular momentum &  \ref{060831a} \\
$ m $  & Mercury's mass &  \ref{090418a} \\
$ m_\odot $  & solar mass &  \ref{021209a} \\
$ M $ & mean anomaly &  \ref{061120ga} \\
$ n $ & mean motion &  \ref{030123g} \\
$ \vv{N} $ & non-radial inertial pressure torque &  \ref{050514a} \\
$ N (e) $ & tidal eccentricity function &  \ref{030218a} \\
$ p $ & semi-integer indicating the resonance &  \ref{061120ga} \\
$ \vv{p} $ & unit vector for averaged precession &  \ref{061115a} \\
$ \vv{P} $ & precession torque &  \ref{V12} \\
$ P_\mathrm{cap}^\pm $ & probability of capture into resonance &  \ref{021024d} \\
$ P_\kk $ & projection of $ \vv{P} $ over $ \vv{\kk} $  &  \ref{050428b} \\
$ P_l $ & Legendre polynomials of degree $l$ &  \ref{090418a} \\
$ P_q $ & projection of $ \vv{P} $ over $ \vv{q} $  &  \ref{061115e} \\
$ P (t) $ & function of the precession torque $ P_\kk $  &  \ref{061116c} \\
$ p $ & semi-integer for core-mantle friction &  \ref{061122v} \\
$ \vv{q} $ & unit vector normal to averaged precession &  \ref{061115c} \\
\hline \end{tabular} 
\end{table}

\begin{table}
\begin{tabular}{|c|l|c|} \hline
 {\it Symbol}  & $ 
\quad \quad \quad \quad \quad $ Designation $ 
\quad \quad \quad \quad \quad \quad $  & Eq. \\  \hline
$ Q $ & quality factor & \ref{090419e} \\
$ \vv{r} $ & radial distance from Mercury's center &  \ref{090418a} \\
$ \vv{r}' $ & radial distance from Mercury's center &  \ref{090419b} \\
$ \vv{\hat r} $ & unit vector for $\vv{r}$ &  \ref{090418a} \\
$ R $ & Mercury's radius &  \ref{050109a}  \\
$ R_c $ & Mercury's core radius &  \ref{V14a}  \\
$ s_g $ & signal function $ s_g = \mathrm{\omega} $ &  \ref{040811r} \\
$ S $ & angle between two directions &  \ref{090419b} \\
$ \vv{T} $ & tidal torque &  \ref{V12} \\
$ T_\kk $ & projection of $ \vv{T} $ over $ \vv{\kk} $  &  \ref{050428b} \\
$ T_q $ & projection of $ \vv{T} $ over $ \vv{q} $  &  \ref{061115e} \\
$ {\cal U} $ & potential energy  &  \ref{021209a} \\
$ {\cal U}_r $ & resonant part of the potential energy  &  \ref{050602z} \\
$ \overline{\cal U} $ & averaged potential energy  &  \ref{030804a} \\
$ v $ & true anomaly &  \ref{090418b}  \\
$ {\cal V} $ & gravitational potential &  \ref{090418a} \\
$ {\cal V}' $ & scalar potential raising tides &  \ref{090419b} \\
$ {\cal V}^g $ & tidal potential &  \ref{021010a} \\
$ w $ & true longitude of date & \ref{090418c} \\
$ x $ & cosine of the obliquity ($x=\cos \ve$) &  \ref{030804a} \\
$ X $ & projection of $ \vv{L} $ on the ecliptic's normal &  \ref{021014c} \\
$ \alpha $ & precession constant &  \ref{061120a} \\
$ \alpha_r $ & libration amplitude &  \ref{090419a} \\
$ \beta $ & libration amplitude &  \ref{030123g} \\
$ \beta_m $ & libration amplitude ($\beta_m = \beta_x / c_m$) &  \ref{021016z} \\
$ \beta_x $ & libration amplitude &  \ref{030731b} \\
$ \gamma $  &  relative rotation angle &  \ref{061129g} \\
$ \vv{\Gamma} $ & efective friction torque &  \ref{020410a} \\
$ \vv{\delta} $ & differential core rotation rate &  \ref{050514a} \\
$ \delta_\kk $ & projection of $ \vv{\delta} $ over $ \vv{\kk} $  & \ref{061115d} \\
$ \delta_p $ & projection of $ \vv{\delta} $ over $ \vv{p} $  &  \ref{061115d} \\
$ \delta (\sigma) $ & tidal phase lag &  \ref{090419d} \\
$ \delta E_d $ & residual dynamical ellipticity &  \ref{050109a} \\
$ \Delta E $ & tidal energy dissipated per cycle &  \ref{090419e} \\
$ \Delta t $ & tidal time lag &  \ref{090419d} \\
$ \Delta \omega $ & width of the resonance &  \ref{021016x} \\
$ \ve $  & obliquity &  \ref{090418c}  \\
$ \ve_i $  & initial obliquity &  \ref{eqANS}  \\
$ \zeta $ & dimensionless parameter &  \ref{090423a} \\
$ \theta $ & hour angle &  \ref{090418c} \\
$ \Theta_\sigma^L $ & tidal coefficient &  \ref{021010c1} \\
$ \Theta_\sigma^\ve $ & tidal coefficient &  \ref{021010c2} \\
$ \kappa $ & effective coupling parameter &  \ref{020410a} \\
$ \kappa_\mathrm{vis} $ & viscous coupling parameter &  \ref{V14a} \\
$ \kappa' $ & effective coupling parameter &  \ref{020410a} \\
$ \kappa_\mathrm{vis}' $ & viscous coupling parameter &  \ref{V14a} \\
$ \kappa_m $ & effective coupling parameter ($ \kappa_m = \kappa / c_m $) & \ref{061115z} \\
$ \mu $ & dimensionless parameter &  \ref{061204i} \\
$ \mu_e $ & body rigidity &  \ref{090419g} \\
$ \mu_\zeta $ & dimensionless parameter &  \ref{081222a} \\
$ \nu $ & kinematic viscosity &  \ref{V14a} \\
$ \nu_e $ & body viscosity &  \ref{090419g} \\
$ \xi $ & internal structure factor &  \ref{eq3} \\
$ \varpi $ & longitude of the perihelion &  \ref{090418b}  \\
\hline \end{tabular} 
\end{table}

\begin{table}
\begin{tabular}{|c|l|c|} \hline
 {\it Symbol}  & $ 
\quad \quad \quad \quad \quad $ Designation $ 
\quad \quad \quad \quad \quad \quad $  & Eq. \\  \hline
$ \rho $ & Mercury's mean density  &  \ref{090419g}  \\
$ \varrho $ & dimensionless parameter &  \ref{040820i} \\
$ \sigma $ & tidal frequency &  \ref{090419d} \\
$ \tau_a $ & time constant for damping body tides &  \ref{090419f} \\
$ \tau_b $ & time constant for damping body tides &  \ref{090419f} \\
$ \phi $ &  libration phase $ (\phi = \varpi + \psi) $ &  \ref{030804a} \\
$ \phi_r $ &  libration phase  &  \ref{090419a} \\
$ \phi_x $ &  libration phase  &  \ref{030731c} \\
$ \psi $ & general precession angle &  \ref{021014c} \\
$ \dot \psi_x $ & general precession ($\dot \psi_x = \dot \phi_x + \dot \psi \cos \ve $) &  \ref{021016z} \\
$ \vv{\omega} $  & mantle's rotation rate &  \ref{060831a} \\
$ \vv{\omega}_c $  & core's rotation rate &  \ref{050514a} \\
$ \omega_c^\kk $ & projection of $ \vv{\omega}_c $ over $ \vv{\kk} $ & \ref{061114a}  \\
$ \omega_i $ & initial rotation rate &  \ref{eqANS} \\
$ \omega_e $ & equilibrium rotation rate &  \ref{040812c} \\
$ \vv{\Omega} $  & precession angular velocity  &  \ref{060831a} \\
$ \Omega (e) $ & tidal eccentricity function &  \ref{030218b} \\
\hline \end{tabular} 
\end{table}

\bibliographystyle{elsarticle-harv}      
\bibliography{correia}

\begin{thebibliography}{57}
\expandafter\ifx\csname natexlab\endcsname\relax\def\natexlab#1{#1}\fi
\expandafter\ifx\csname url\endcsname\relax
  \def\url#1{\texttt{#1}}\fi
\expandafter\ifx\csname urlprefix\endcsname\relax\def\urlprefix{URL }\fi

\bibitem[{{Anderson} et~al.(1987){Anderson}, {Colombo}, {Espsitio}, {Lau}, and
  {Trager}}]{Anderson_etal_1987}
{Anderson}, J.~D., {Colombo}, G., {Espsitio}, P.~B., {Lau}, E.~L., {Trager},
  G.~B., Sep. 1987. {The mass, gravity field, and ephemeris of Mercury}. Icarus
  71, 337--349.

\bibitem[{{Andoyer}(1923)}]{Andoyer_1923}
{Andoyer}, H., Mar. 1923. {Cours de M\'{e}canique C\'{e}leste}.
  Gauthier-Villars, Paris.

\bibitem[{{Burns}(1976)}]{Burns_1976}
{Burns}, J.~A., Aug. 1976. {Consequences of the tidal slowing of Mercury}.
  Icarus 28, 453--458.

\bibitem[{{Colombo}(1965)}]{Colombo_1965}
{Colombo}, G., 1965. {Rotational Period of the Planet Mercury}. \nat 208,
  575--578.

\bibitem[{{Colombo} and {Shapiro}(1966)}]{Colombo_Shapiro_1966}
{Colombo}, G., {Shapiro}, I.~I., Jul. 1966. {The Rotation of the Planet
  Mercury}. \apj 145, 296--307.

\bibitem[{{Correia}(2006)}]{Correia_2006}
{Correia}, A.~C.~M., Dec. 2006. {The core-mantle friction effect on the secular
  spin evolution of terrestrial planets}. \epsl 252, 398--412.

\bibitem[{{Correia} and {Laskar}(2001)}]{Correia_Laskar_2001}
{Correia}, A.~C.~M., {Laskar}, J., Jun. 2001. {The four final rotation states
  of Venus}. \nat 411, 767--770.

\bibitem[{{Correia} and {Laskar}(2003)}]{Correia_Laskar_2003I}
{Correia}, A.~C.~M., {Laskar}, J., May 2003. {Long-term evolution of the spin
  of Venus II. Numerical simulations}. Icarus 163, 24--45.

\bibitem[{{Correia} and {Laskar}(2004)}]{Correia_Laskar_2004}
{Correia}, A.~C.~M., {Laskar}, J., Jun. 2004. {Mercury's capture into the 3/2
  spin-orbit resonance as a result of its chaotic dynamics}. \nat 429,
  848--850.

\bibitem[{{Correia} and {Laskar}(2009)}]{Correia_Laskar_2009}
{Correia}, A.~C.~M., {Laskar}, J., May 2009. {Mercury's capture into the 3/2
  spin-orbit resonance including the effect of core-mantle friction}. Icarus
  201, 1--11.

\bibitem[{{Correia} et~al.(2003){Correia}, {Laskar}, and {N\'eron de
  Surgy}}]{Correia_etal_2003}
{Correia}, A.~C.~M., {Laskar}, J., {N\'eron de Surgy}, O., May 2003. {Long-term
  evolution of the spin of Venus I. Theory}. Icarus 163, 1--23.

\bibitem[{{Counselman} and {Shapiro}(1970)}]{Counselman_Shapiro_1970}
{Counselman}, C.~C., {Shapiro}, I.~I., 1970. {Spin-Orbit resonance of Mercury}.
  Symposia Mathematica 3, 121--169.

\bibitem[{{Darwin}(1880)}]{Darwin_1880}
{Darwin}, G.~H., 1880. {On the secular change in the elements of a satellite
  revolving around a tidally distorted planet}. Philos. Trans. R. Soc. London
  171, 713--891.

\bibitem[{{Darwin}(1908)}]{Darwin_1908}
{Darwin}, G.~H., 1908. {Scientific Papers}. Cambridge University Press.

\bibitem[{{Defrancesco}(1988)}]{Defrancesco_1988}
{Defrancesco}, S., Apr. 1988. {Schiaparelli's determination of the rotation
  period of Mercury: a re-examination}. Journal of the British Astronomical
  Association 98, 146--150.

\bibitem[{{Deleplace} and {Cardin}(2006)}]{Deleplace_Cardin_2006}
{Deleplace}, B., {Cardin}, P., Nov. 2006. {Viscomagnetic torque at the core
  mantle boundary}. \gji 167, 557--566.

\bibitem[{{Dones} and {Tremaine}(1993)}]{Dones_Tremaine_1993}
{Dones}, L., {Tremaine}, S., May 1993. {On the origin of planetary spins}.
  Icarus 103, 67--92.

\bibitem[{{Gans}(1972)}]{Gans_1972}
{Gans}, R.~F., 1972. {Viscosity of the Earth's core}. \jgr 77, 360--366.

\bibitem[{{Goldreich}(1966)}]{Goldreich_1966}
{Goldreich}, P., Feb. 1966. {Final spin states of planets and satellites}. \aj
  71, 1--7.

\bibitem[{{Goldreich} and {Peale}(1966)}]{Goldreich_Peale_1966}
{Goldreich}, P., {Peale}, S., Aug. 1966. {Spin-orbit coupling in the solar
  system}. \aj 71, 425--438.

\bibitem[{{Goldreich} and {Peale}(1967)}]{Goldreich_Peale_1967}
{Goldreich}, P., {Peale}, S., Jun. 1967. {Spin-orbit coupling in the solar
  system. II. The resonant rotation of Venus}. \aj 72, 662--668.

\bibitem[{{Henrard}(1993)}]{Henrard_1993}
{Henrard}, J., 1993. {The adiabatic invariant in classical dynamics}. In:
  Dynamics Reported. Springer Verlag, New York, pp. 117--235.

\bibitem[{{Hut}(1981)}]{Hut_1981}
{Hut}, P., Jun. 1981. {Tidal evolution in close binary systems}. \aap 99,
  126--140.

\bibitem[{{Kaula}(1964)}]{Kaula_1964}
{Kaula}, W.~M., 1964. {Tidal dissipation by solid friction and the resulting
  orbital evolution}. \rg 2, 661--685.

\bibitem[{{Kinoshita}(1977)}]{Kinoshita_1977}
{Kinoshita}, H., Apr. 1977. {Theory of the rotation of the rigid earth}.
  Celestial Mechanics 15, 277--326.

\bibitem[{{Kokubo} and {Ida}(2007)}]{Kokubo_Ida_2007}
{Kokubo}, E., {Ida}, S., Dec. 2007. {Formation of Terrestrial Planets from
  Protoplanets. II. Statistics of Planetary Spin}. \apj 671, 2082--2090.

\bibitem[{{Lambeck}(1980)}]{Lambeck_1980}
{Lambeck}, K., 1980. {The Earth's Variable Rotation: Geophysical Causes and
  Consequences}. Cambridge University Press.

\bibitem[{{Laskar} and {Robutel}(1993)}]{Laskar_Robutel_1993}
{Laskar}, J., {Robutel}, P., Feb. 1993. {The chaotic obliquity of the planets}.
  \nat 361, 608--612.

\bibitem[{{Levrard} et~al.(2007){Levrard}, {Correia}, {Chabrier}, {Baraffe},
  {Selsis}, and {Laskar}}]{Levrard_etal_2007}
{Levrard}, B., {Correia}, A.~C.~M., {Chabrier}, G., {Baraffe}, I., {Selsis},
  F., {Laskar}, J., Jan. 2007. {Tidal dissipation within hot Jupiters: a new
  appraisal}. \aap 462, L5--L8.

\bibitem[{{Lumb} and {Aldridge}(1991)}]{Lumb_Aldridge_1991}
{Lumb}, L.~I., {Aldridge}, K.~D., 1991. {On viscosity estimates for the Earth's
  fluid outer core-mantle coupling}. J. Geophys. Geoelectr. 43, 93--110.

\bibitem[{{Margot} et~al.(2007){Margot}, {Peale}, {Jurgens}, {Slade}, and
  {Holin}}]{Margot_etal_2007}
{Margot}, J.~L., {Peale}, S.~J., {Jurgens}, R.~F., {Slade}, M.~A., {Holin},
  I.~V., May 2007. {Large Longitude Libration of Mercury Reveals a Molten
  Core}. Science 316, 710--714.

\bibitem[{{Mathews} and {Guo}(2005)}]{Mathews_Guo_2005}
{Mathews}, P.~M., {Guo}, J.~Y., Feb. 2005. {Viscoelectromagnetic coupling in
  precession-nutation theory}. \jgr (Solid Earth) 110, B02402--16.

\bibitem[{{McGovern} et~al.(1965){McGovern}, {Gross}, and
  {Rasool}}]{McGovern_etal_1965}
{McGovern}, W.~E., {Gross}, S.~H., {Rasool}, S.~I., 1965. {Rotation period of
  the planet Mercury}. \nat 208, 375.

\bibitem[{{Mignard}(1979)}]{Mignard_1979}
{Mignard}, F., May 1979. {The evolution of the lunar orbit revisited. I}. Moon
  and Planets 20, 301--315.

\bibitem[{{Mignard}(1980)}]{Mignard_1980}
{Mignard}, F., Oct. 1980. {The evolution of the lunar orbit revisited. II}.
  Moon and Planets 23, 185--201.

\bibitem[{{Munk} and {MacDonald}(1960)}]{Munk_MacDonald_1960}
{Munk}, W.~H., {MacDonald}, G.~J.~F., 1960. {The Rotation of the Earth; A
  Geophysical Discussion}. Cambridge University Press.

\bibitem[{{Murdin}(2000)}]{Murdin_2000}
{Murdin}, P., Nov. 2000. {Caloris Basin}. Encyclopedia of Astronomy and
  Astrophysics.

\bibitem[{{Ness}(1978)}]{Ness_1978}
{Ness}, N.~F., Mar. 1978. {Mercury - Magnetic field and interior}. Space
  Science Reviews 21, 527--553.

\bibitem[{{Ness} et~al.(1975){Ness}, {Behannon}, {Lepping}, and
  {Whang}}]{Ness_etal_1975}
{Ness}, N.~F., {Behannon}, K.~W., {Lepping}, R.~P., {Whang}, Y.~C., Jul. 1975.
  {The magnetic field of Mercury. I}. \jgr 80, 2708--2716.

\bibitem[{{Ness} et~al.(1974){Ness}, {Behannon}, {Lepping}, {Whang}, and
  {Schatten}}]{Ness_etal_1974}
{Ness}, N.~F., {Behannon}, K.~W., {Lepping}, R.~P., {Whang}, Y.~C., {Schatten},
  K.~H., Jul. 1974. {Magnetic field observations near Mercury: Preliminary
  results from Mariner 10}. Science 185, 153--162.

\bibitem[{{Noir} et~al.(2003){Noir}, {Cardin}, {Jault}, and
  {Masson}}]{Noir_etal_2003}
{Noir}, J., {Cardin}, P., {Jault}, D., {Masson}, J.-P., Aug. 2003.
  {Experimental evidence of non-linear resonance effects between retrograde
  precession and the tilt-over mode within a spheroid}. \gji 154, 407--416.

\bibitem[{{Pais} et~al.(1999){Pais}, {Le Mou{\"e}l}, {Lambeck}, and
  {Poirier}}]{Pais_etal_1999}
{Pais}, M.~A., {Le Mou{\"e}l}, J.~L., {Lambeck}, K., {Poirier}, J.~P., Dec.
  1999. {Late Precambrian paradoxical glaciation and obliquity of the Earth - a
  discussion of dynamical constraints}. \epsl 174, 155--171.

\bibitem[{{Peale}(1974)}]{Peale_1974}
{Peale}, S.~J., Jun. 1974. {Possible histories of the obliquity of Mercury}.
  \aj 79, 722--744.

\bibitem[{{Peale}(1976)}]{Peale_1976}
{Peale}, S.~J., Aug. 1976. {Inferences from the dynamical history of Mercury's
  rotation}. Icarus 28, 459--467.

\bibitem[{{Peale} and {Boss}(1977)}]{Peale_Boss_1977}
{Peale}, S.~J., {Boss}, A.~P., Aug. 1977. {Spin-orbit constraint on the
  viscosity of a Mercurian liquid core}. \jgr 82, 743--749.

\bibitem[{{Pettengill} and {Dyce}(1965)}]{Pettengill_Dyce_1965}
{Pettengill}, G.~H., {Dyce}, R.~B., 1965. {A Radar Determination of the
  Rotation of the Planet Mercury}. \nat 206, 1240--1241.

\bibitem[{{Poincar\'e}(1910)}]{Poincare_1910}
{Poincar\'e}, H., 1910. {Sur la pr\'{e}cession des corps d\'{e}formables}.
  Bull. Astron. 27, 321--356.

\bibitem[{{Poirier}(1988)}]{Poirier_1988}
{Poirier}, J.~P., Jan. 1988. {Transport properties of liquid metals and
  viscosity of the earth's core}. Geophysical Journal 92, 99--105.

\bibitem[{{Rochester}(1976)}]{Rochester_1976}
{Rochester}, M.~G., 1976. {The secular decrease of obliquity due to dissipative
  core-mantle coupling}. Geophys. J.R.A.S. 46, 109--126.

\bibitem[{{Sasao} et~al.(1980){Sasao}, {Okubo}, and {Saito}}]{Sasao_etal_1980}
{Sasao}, T., {Okubo}, S., {Saito}, M., 1980. {A Simple Theory on Dynamical
  Effects of Stratified Fluid Core upon Nutational Motion of the Earth}. In:
  IAU Symp.78: Nutation and the Earth's Rotation. pp. 165--183.

\bibitem[{{Schiaparelli}(1890)}]{Schiaparelli_1890}
{Schiaparelli}, G.~V., 1890. {Sulla rotazione di Mercurio}. Astronomische
  Nachrichten 123, 241--250.

\bibitem[{{Smart}(1953)}]{Smart_1953}
{Smart}, W.~M., 1953. {Celestial Mechanics.} London, New York, Longmans, Green.

\bibitem[{{Spohn} et~al.(2001){Spohn}, {Sohl}, {Wieczerkowski}, and
  {Conzelmann}}]{Spohn_etal_2001}
{Spohn}, T., {Sohl}, F., {Wieczerkowski}, K., {Conzelmann}, V., Dec. 2001. {The
  interior structure of Mercury: what we know, what we expect from
  BepiColombo}. \planss 49, 1561--1570.

\bibitem[{{Strom} et~al.(2008){Strom}, {Chapman}, {Merline}, {Solomon}, and
  {Head}}]{Strom_etal_2008}
{Strom}, R.~G., {Chapman}, C.~R., {Merline}, W.~J., {Solomon}, S.~C., {Head},
  J.~W., Jul. 2008. {Mercury Cratering Record Viewed from MESSENGER's First
  Flyby}. Science 321, 79--81.

\bibitem[{{Tisserand}(1891)}]{Tisserand_1891}
{Tisserand}, F., 1891. {Trait\'e de M\'ecanique C\'eleste (Tome II)}.
  Gauthier-Villars, Paris.

\bibitem[{{Wijs} et~al.(1998){Wijs}, {Kresse}, {Vo\v{c}adlo}, {Dobson}, {Alfe},
  {Gillan}, and {Price}}]{deWijs_etal_1998}
{Wijs}, G.~A., {Kresse}, G., {Vo\v{c}adlo}, L., {Dobson}, D., {Alfe}, D.,
  {Gillan}, M.~J., {Price}, G.~D., 1998. {The viscosity of liquid iron at the
  physical conditions of the Earth's core}. \nat 392, 805--807.

\bibitem[{{Yoder}(1995)}]{Yoder_1995}
{Yoder}, C.~F., Oct. 1995. {Venus' free obliquity.} Icarus 117, 250--286.

\end{thebibliography}

\prep{\end{document}}

\setcounter{figure}{0}
\setcounter{table}{0}

\clearpage
\section*{Tables and Figures}


\TabA
\clearpage

\TabB
\clearpage

\TabC
\clearpage


\FigA
\clearpage

\FigB
\clearpage

\FigC
\clearpage

\FigD
\clearpage

\FigE
\clearpage

\FigF
\clearpage


\FigH
\clearpage

\FigI

\FigJ
\clearpage

\FigK
\clearpage


\end{document}